\documentclass[12pt, onecolumn, letterpaper]{article}
\usepackage[
	letterpaper, 
	left=0.75in,
	right=0.75in,
	top=0.75in,
	bottom=0.75in
]{geometry} 

\usepackage[dvipsnames]{xcolor}

\usepackage[
	colorlinks=true,
	allcolors=black,
	linkcolor=blue,
	urlcolor=blue,
	citecolor=blue,
]{hyperref} 
\usepackage{caption}
\usepackage{pgf}
\usepackage{graphicx}
\usepackage{tabularx, booktabs, multirow}
\usepackage{amsmath, amsthm, amsfonts, amssymb, bbm}

\usepackage{lineno}
\usepackage{setspace}

\usepackage[affil-it]{authblk}
\usepackage[super, comma, sort&compress]{natbib}
\title{\textbf{\textsf{Encoding protein dynamic information in graph representation for functional residue identification}}}
\author[1]{Yuan Chiang}
\author[1]{Wei-Han Hui}
\author[1, 2]{Shu-Wei Chang\thanks{Corresponding author: \href{mailto:changsw@ntu.edu.tw}{changsw@ntu.edu.tw}}\thanks{Lead contact}}
\affil[1]{\normalsize Department of Civil Engineering, National Taiwan University, Taipei 10663, Taiwan}
\affil[2]{\normalsize Department of Biomedical Engineering, National Taiwan University, Taipei 10663, Taiwan}
\date{\today}

\begin{document}
\maketitle
\sloppy
\doublespacing

{
\sf
Recent advances in protein function prediction exploit graph-based deep learning approaches to correlate the structural and topological features of proteins with their molecular functions. However, proteins \textit{in vivo} are not static but dynamic molecules that alter conformation for functional purposes. Here we apply normal mode analysis to native protein conformations and augment protein graphs by connecting edges between dynamically correlated residue pairs. In the multilabel function classification task, our method demonstrates a remarkable performance gain based on this dynamics-informed representation. The proposed graph neural network, ProDAR, increases the interpretability and generalizability of residue-level annotations and robustly reflects structural nuance in proteins. We elucidate the importance of dynamic information in graph representation by comparing class activation maps for hMTH1, nitrophorin, and SARS-CoV-2 receptor binding domain. Our model successfully learns the dynamic fingerprints of proteins and pinpoints the residues of functional impacts, with vast untapped potential for broad biotechnology and pharmaceutical applications.

\bigskip

}

\clearpage

\begin{figure*}[htb!]
    \centering
    \includegraphics[width=\textwidth]{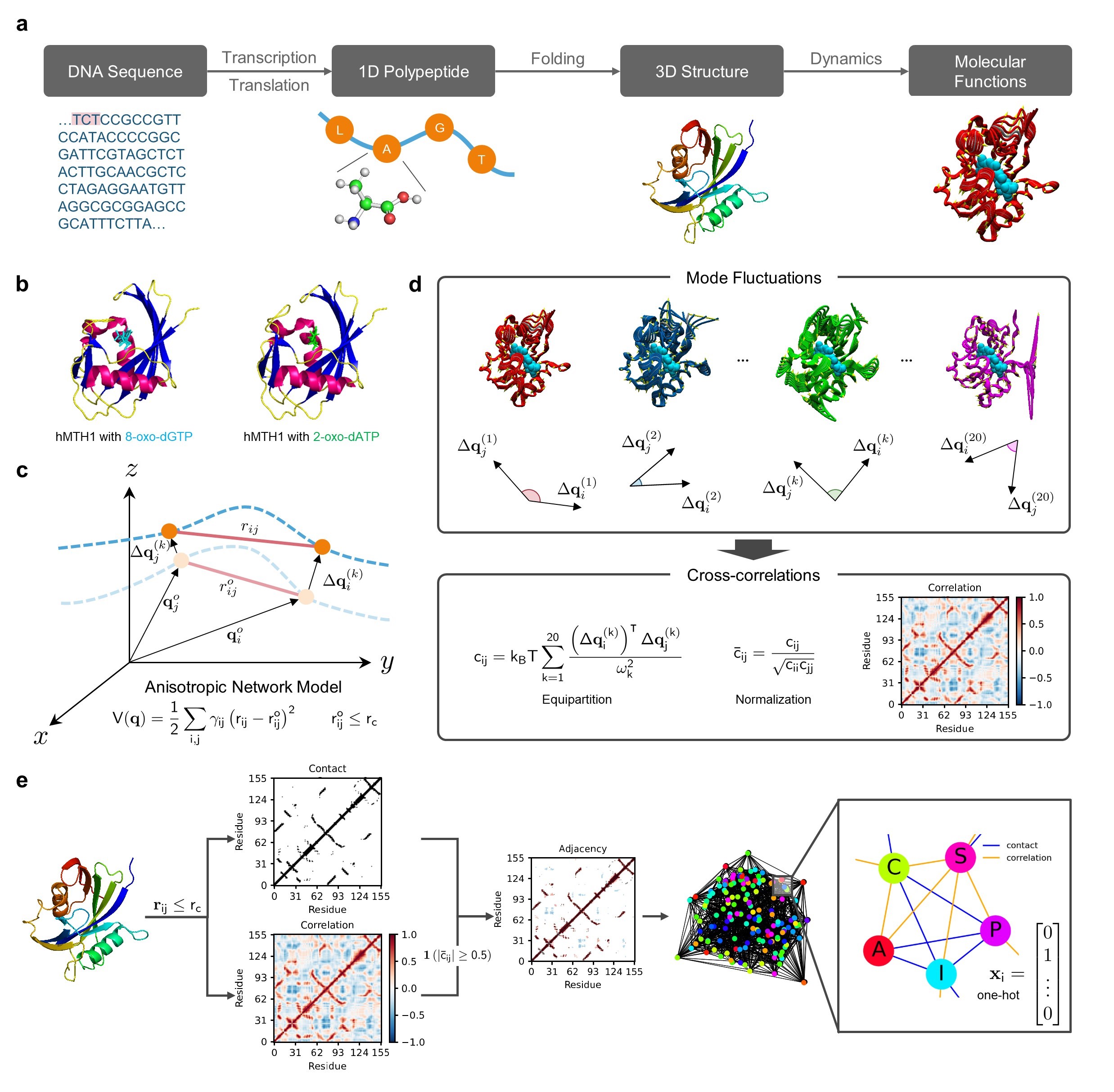}
    \caption{\textbf{Protein dynamics in relation to molecular functions and graph encoding of dynamic information. a} Protein hierarchy from DNA sequence, to polypeptide chain, to folding structure, to molecular functions. \textbf{b} Conformational difference between hMTH1 complexes with 8-oxo-dGTP and 2-oxo-dATP. The binding pockets in the two complexes have different hydrogen bonding patterns with substrates. \textbf{c} Equilibrium fluctuation about the native conformation and the energy induced by conformational change. \textbf{d} Twenty mode fluctuations and the cross-correlations between residue pairs. The correlation map is calculated by the trace of cross-correlation (inner product of the fluctuation vectors), followed by normalization against the diagonal components. \textbf{e} Graph construction scheme used in this work, merging the contact map and correlation map into an adjacency matrix for graph representation. One-hot vectors of 21 dimensions are used to encode different types of amino acids.}
    \label{fig:1}
\end{figure*}

Proteins are molecular machines carrying out a variety of functions in biological processes. From 1D sequences to 3D structures, the genetic codes determine the sequence of amino acids and the way proteins fold into 3D structures (Figure \ref{fig:1}a). With great experimental efforts, \textit{e.g.} with X-ray crystallography (XRC) and cryogenic electron microscopy (cryo-EM), many protein structures have been determined at high resolutions, typically spanning from 1 to 4 \AA\cite{drenth2007principles, bai2015cryo, danev2019cryo, matsumoto2021extraction}. To date, over 170,000 entries of 3D protein structures can be found in the Protein Data Bank (PDB)\cite{berman2003announcing}, the repository of experimentally determined 3D structures of proteins, nucleic acids, and complex assemblies. However, the number of available protein structures falls far short of the available sequence data. For example, UniProt Knowledge-base (UniProtKB)\cite{uniprot2021uniprot}, the database of protein sequence and function, contains over 200 million annotated sequences.

To fill the gap between sequence and structure data and hence explore the unknown functions of newly discovered proteins, the determination of protein structures by experimental and computational techniques has been a long-standing issue in structural biology. Homology-modeling techniques have driven advances in the comparative construction and evaluation of protein structures\cite{webb2016comparative, waterhouse2018swiss}. Comparative construction means such homology-modeling techniques need \textit{a priori} structures as templates to build the models, which inevitably limits the generalizability toward the large proportion of completely unknown structures. The recent development of deep learning methods takes advantage of an increasing amount of sequence data to predict 3D protein structure. In the 14th biennial Critical Assessment of Structure Prediction (CASP14), deep learning models AlphaFold2\cite{jumper2021highly} and RoseTTAFold\cite{baek2021accurate} achieved supremacy in the accuracy of blind structure prediction tests. At the frontier of deep learning approaches, both models took input from multiple sequence alignments (MSA) and applied attention-based message passing schemes on residue pairs. In particular, their integration with roto-translationally equivalent SE(3)-transformer and invariant point attention\cite{fuchs2020se, ingraham2019generative} plays a crucial role in the fine adjustment of torsional angles in backbones and side chains. With the aid of the high-throughput structures predicted by these models, we can now tackle the next challenge in protein function prediction by data-driven approaches.

In recent decades, there has been an emerging paradigm that, in proteins, \emph{sequence-encodes-structure-encodes-dynamics-encodes-function}\cite{bahar2010normal} (Figure \ref{fig:1}a). With the structures determined, the dynamics in turn regulate the behaviors and functions of proteins. For example, human MutT homolog 1 (hMTH1), as shown in Figure \ref{fig:1}b, utilizes different conformations of the same substrate binding pocket to recognize different types of oxidized nucleotides (8-oxo-dGTP and 2-oxo-dATP) and to conduct nucleotide sanitization\cite{waz2017structural}. In many cases, protein functions are intertwined with dynamic properties ranging from local residue fluctuations to global collective motions. More essentially, the functional prediction of proteins is a great challenge, and the problems are multifold. Traditional machine learning classifiers, such as support vector machines, random forests, and gradient boosted decision trees, have been extensively used to predict site-specific or protein-level functions\cite{koo2019towards, das2021cath}. Many of the recent deep learning models, on the other hand, focus on the variations in sequence embedding and neural network architecture\cite{gligorijevic2021structure, villegas2021unsupervised, sanyal2020proteingcn, swenson2020persgnn}. One of the most natural choices for learning protein features is the graph neural network (GNN). GNNs can aggregate local residue features in a high-dimensional space and learn appropriate representations for local and global inferences\cite{sanyal2020proteingcn}. Although the dynamical features of proteins had been considered in many computational approaches for function inference and mutation detection\cite{ponzoni2018structural, demir2011ensemble,gheeraert2019exploring}, the systematic way to incorporate dynamical information in deep learning frameworks remains elusive. Even a tiny point mutation in the protein sequence or excess thermal overflow can distort the backbone structure, alter the dynamic behavior, and eventually nullify certain functions as a whole. It is therefore expected that encoding dynamic information of proteins into GNNs would increase the discriminatory capacity of neural networks for either regular function prediction or anomaly (for example, mutation) detection. 

In this work, we propose a framework to incorporate dynamic information into graph-based deep neural networks. By processing protein structures into three feature pipelines, including spatial, topological, and dynamic features, we successfully increase the discriminatory power of neural networks and enhance the classification performance on protein function prediction. We retrieve 3D structures of proteins from PDB\cite{berman2003announcing, mir2018pdbe, burley2021rcsb, kinjo2016protein, kinjo2018new} and perform multilabel classification on molecular function (MF) gene ontology (GO) annotations obtained from Structure Integration with Function, Taxonomy, and Sequence (SIFTS) database\cite{velankar2012sifts, mir2018pdbe, dana2019sifts}. We also use the gradient-weighted class activation map (Grad-CAM)\cite{selvaraju2017grad} to identify the dynamically activated residues (DARs) by comparing the activation maps with and without dynamic information. Our model, protein dynamically activated residue (ProDAR), determines the residues that are critical to protein dynamic behaviors and that have fundamental importance for protein and drug engineering. We also demonstrate several examples of DARs identified by ProDAR and discuss their implications on the dynamic characteristics of protein functions.

\clearpage

\section*{Results}
\subsection*{ProDAR encodes the dynamic information of proteins in graph representation}

Proteins \textit{in vivo} are neither static nor rigid bodies but dynamic and deformable molecules undergoing conformational changes due to thermal fluctuations. Via exploration by thermodynamical process, particular atomistic configurations are sampled in large conformational space to obtain the desired behaviors. To screen thousands of conformational changes in a high-throughput manner, we apply normal mode analysis (NMA) to the anisotropic network model (ANM), a surrogate model where molecular structures are represented as bead-spring networks. As illustrated in Figure \ref{fig:1}c, $\textrm{C}^\alpha$ atoms are extracted from the polypeptide chain as beads, and any two beads within a prescribed cutoff distance $r_c$ are considered in contact and are connected with each other through a spring. The side-chain-side-chain contacts are ignored based on the assumption that the side-chain interactions implicitly determine the distances between $\textrm{C}^\alpha$ atoms and therefore the contact maps. The formed bead-spring network has zero potential energy at the native conformation, and the energy incurred by conformation change is then the sum of spring harmonic potential (see Methods). We collect the first 20 mode fluctuations (mode shapes) obtained from the eigenanalysis of the potential Hessian matrix and compute the cross-correlations between all residue pairs (Figure \ref{fig:1}d). The cross-correlations underscore the residue pairs that have a high degree of synergy across 20 mode fluctuations. The pairs that mostly move toward the same direction have high correlation, while that mostly move toward the opposite directions have high anticorrelation. For each protein, we calculate the correlation map by taking the trace of the cross-correlation matrix and normalizing along diagonal components. The correlation map thus identifies the residue pairs that contribute to the collective dynamics of protein accessible at the native conformation. 

To encode structural and dynamic information into the protein graph representation, we merge the contact map and correlation map into a single adjacency matrix. The residue pairs in contact are first connected together with \textit{contact edges}. Aside from the residue pairs with contact edges, other residue pairs with absolute correlation values no less than 0.5 ($\bar{c}_{ij} \geq 0.5$) are connected with \textit{correlation edges}. Each node is assigned with a one-hot feature vector of 21 dimensions according to the type of amino acid it holds. The constructed protein graphs therefore encompass sequential, structural, and dynamic information of proteins in a unified representation. 

\subsection*{Correlation edges representing the dynamic features of proteins facilitate message propagation between residues in GNN}

\begin{figure*}[htb!]
    \centering
    \includegraphics[width=\textwidth]{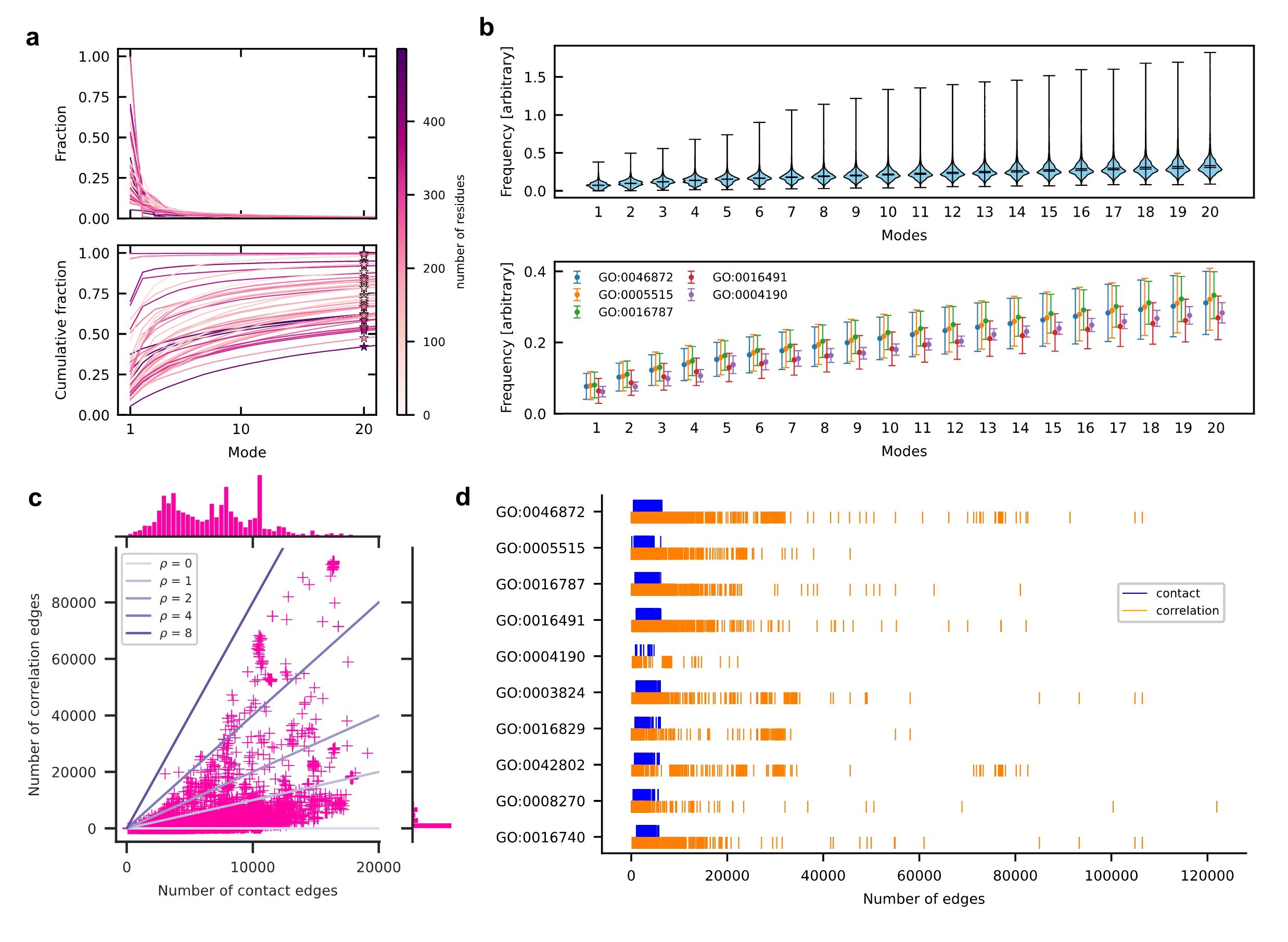}
    \caption{\textbf{Summary of the modal frequencies and correlation edges of proteins. a} Individual and cumulative fractions of the first 20 modes compared to the complete set of modes. 100 protein entries are randomly picked from the dataset. The number of residues are colored accordingly. \textbf{b} Modal frequencies of PDB entries in the dataset. The upper panel shows the distribution of all PDB entries. The bottom panel shows the mean and standard deviation values with respect to the five largest populations of MF-GO annotations. \textbf{c} Number of contact and correlation edges in the dataset. $\rho$ denotes the number ratio of correlation edges to contact edges. \textbf{d} Number of contact and correlation edges for the ten largest populations of MF-GO annotations.}
    \label{fig:2}
\end{figure*}

To explore the dynamic features obtained from NMA, we examine the individual and cumulative fractions of the first 20 modes of randomly selected PDB chains, covering different molecular functions and residue numbers (Figure \ref{fig:2}a). It has been examined by principal component analysis (PCA) that 20 modes are typically sufficient enough to capture the essential dynamics of proteins, even for proteins of heavy molecular weight\cite{david2014principal}. Our results show that the individual contributions from modes decrease drastically as the mode number increases. For the first 20 modes, in these examples, most of the cumulative fractions account for more than 50\% of all available modes (see Methods). The high-frequency modes are relatively negligible compared with low-frequency modes in terms of the global dynamic behavior of proteins. Our method includes modes from low to high frequencies and therefore incorporates the most prominent fluctuations in global protein dynamics. 

The upper panel of Figure \ref{fig:2}b presents the comprehensive frequency distributions of the first 20 modes in our dataset. The average frequencies linearly increase with mode number from 0.075 to 0.331, and the standard deviations increase from 0.040 to 0.118. By truncating the first 20 modes from NMA, a wide range of modal frequencies and fluctuations are extracted and encoded into the correlation maps. For the five largest populations of MF-GO annotations, including metal ion binding (GO:0046872), protein binding (GO:0005515), hydrolase activity (GO:0016787), oxidoreductase activity (GO:0016491), and aspartic-type endopeptidase activity (GO:0004190), universal frequency increments with respect to mode number are observed, as shown in the bottom panel of Figure \ref{fig:2}b. Although the modal frequencies of these functional annotations overlap, these functions have different frequency averages and deviations. It is possible that the multifunctional nature of proteins gives rise to large frequency overlaps across different functional annotations. 

Through our graph construction framework, a considerable number of correlation edges are added to the protein graphs, as shown in Figure \ref{fig:2}c. Although a large number of the PDB entries have few correlation edges, the number of correlation edges in some PDB entries can be nearly eightfold the number of contact edges. Our work has benefited from these correlation edges, which fundamentally change the way GNNs draw inferences and reason down to the residue level. Figure \ref{fig:2}d shows the distributions of contact and correlation edges with respect to ten MF-GO terms in our dataset. Note that the number of correlation edges for aspartic-type endopeptidase activity (GO:0004190) is markedly smaller than those of the other terms. Moreover, zinc ion binding (GO:0008270) has few PDB entries with extremely large numbers of correlation edges (approximately over 120,000). These correlation edges alter protein graph representations and increase the amount of dynamic information encoded into the graphs. The distributions of contact edges are similar across different terms, but those of correlation edges are remarkably dissimilar from each other, indicating that the dynamic information encoded in correlation edges plays an important role in MF-GO terms and that our graph representations contain more information, better representing and differentiating proteins in high-dimensional space. 

\clearpage

\subsection*{Dynamics-informed representation increases the discriminatory power for function classification}

\begin{figure}[htb!]
	\centering
    \includegraphics[width=0.5\textwidth]{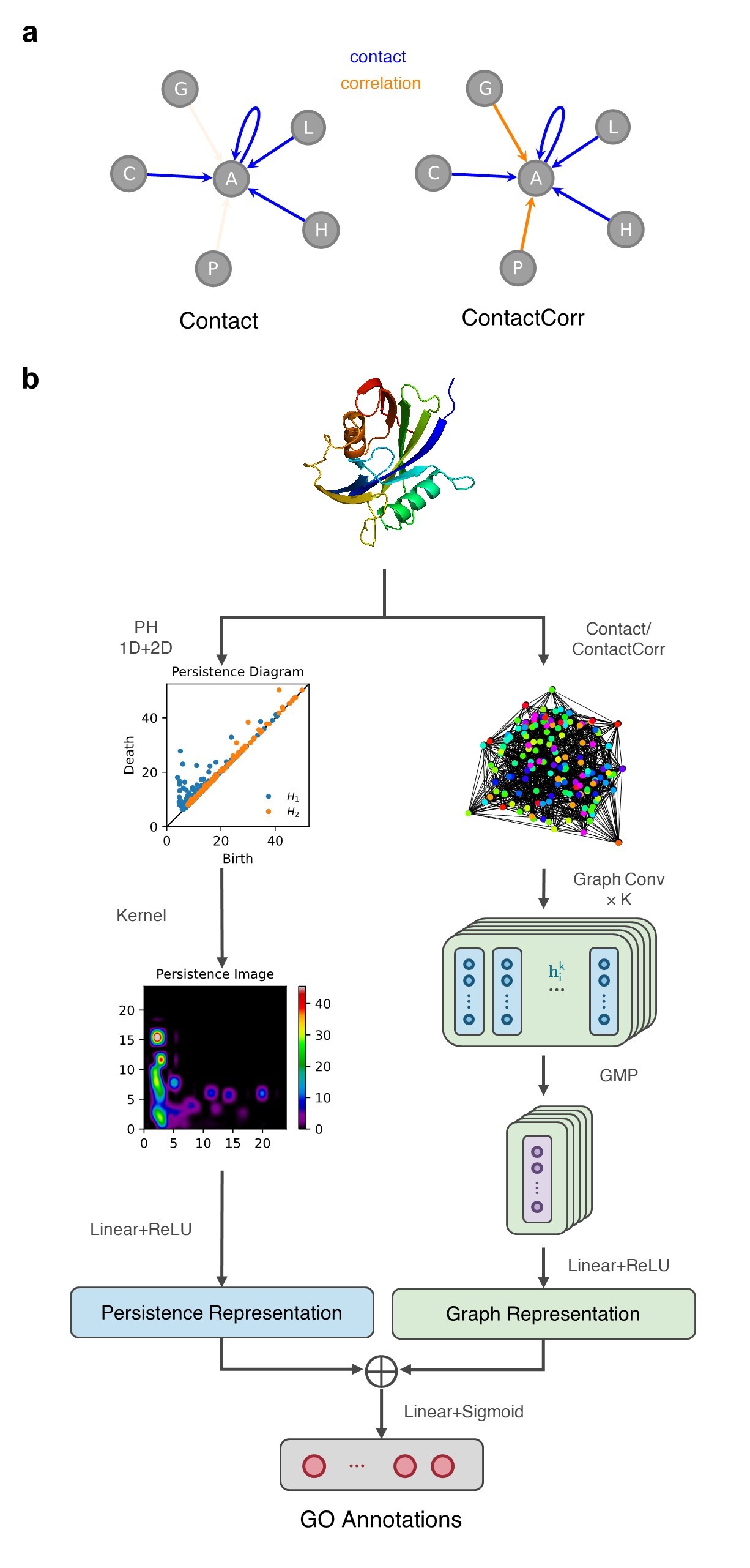}
    \caption{\textbf{ProDAR model architecture. a} Graph with only contact edges (Contact) and graph with both contact and correlation edges (ContactCorr). \textbf{b} Three data pipelines of spatial (contact map), topological (persistence diagram), and dynamic (correlation map) features for learning protein functions. Five graph convolutional layers ($K=5$) and GMP are used to extract the functionally important residues. 1D and 2D persistence diagrams ($H_1, H_2$) are transformed into a persistence image via Gaussian kernels and pixelated integrals. The embedded persistence and graph representations are concatenated into protein representations.}
    \label{fig:3}
\end{figure}

The correlation edges in our graph construction establish connections between residues far apart in space but dynamically correlated in terms of functional motions. These distant, non-local connections ameliorate the information flow in graphs by permitting message to pass over a long distance without the limitation imposed by the depth of graph convolutional layers. This largely alleviates the restriction of the depth of GNNs in consideration of the computing efficiency. To examine the effectiveness of correlation edges for protein function classification, we construct two sets of graphs, one with only contact edges and another with both contact and correlation edges (Figure \ref{fig:3}a). In addition to graphs, we calculate 1D and 2D persistence diagrams of $\textrm{C}^\alpha$ atoms in each protein, the coarsened topological features measured by persistent homology\cite{swenson2020persgnn, edelsbrunner2008persistent}, and embed the diagrams into persistence images via Gaussian kernels and pixelated integrals\cite{adams2017persistence} (see Methods). Our model takes input from three feature pipelines: (1) spatial information from the contact map where residue pairs are closer than cutoff distance $r_c$ such as 8\AA\ and 12\AA, (2) topological information from the persistence image, which is preprocessed as vectors of 625 dimensions, and (3) dynamic information based on the correlation map where residue pairs with absolute correlation no less than 0.5 are connected with correlation edges if they are not in contact. As shown in Figure \ref{fig:3}b, the graphs are fed into five graph convolutional layers in sequence. Each layer of graph convolution takes the adjacency matrix and node features as input and outputs the aggregated node features to the next layer. In this work, we adopt the graph convolutional network (GCN) by Kipf \textit{et al.}\cite{kipf2016semi} and Graph SAmple and aggreGatE (GraphSAGE) by Hamilton \textit{et al}\cite{hamilton2017inductive}. Global maximum pooling (GMP) is applied to readout from five graph convolutional layers the residues that are functionally important and prominent among their nearest to fifth nearest neighbors. Linear layer and rectified linear unit (ReLU) activation are used to embed persistence image and graph pooling layer into vector representations of 512 dimensions. The persistence representation and the graph representation are then concatenated and passed into a fully connected layer with sigmoid activation. The final outputs are the probabilities that the protein of interest pertains to the particular GO annotations (see Methods). 

To examine the utility and importance of structural, topological, and dynamic features in predicting protein functions, we design four model architectures: Contact, ContactPers, ContactCorr, and ContactCorrPers, with each model having different combinations of feature pipelines (Table \ref{table:models}). We also construct two datasets with different cutoff distances $r_c = 8$ \AA\ and $r_c = 12$ \AA\ to observe the effect of the cutoff on the classification performance. In the literature, the choices of cutoff distance for graph-based protein feature extraction vary from 6 to 15 \AA\cite{swenson2020persgnn, gligorijevic2021structure,borgwardt2005protein}. Here we follow the cutoff 8 \AA\ used by PersGNN\cite{swenson2020persgnn} for a fair comparison. Furthermore, to avoid the influence of imbalanced positive and negative samples, we calculate the balanced accuracy (arithmetic mean of the true positive rate and true negative rate) of the test set as the immediate metric of model performance. Due to the usage of weighted binary cross-entropy loss (Methods), the models tend to have higher final recall at the cost of precision. We regard this phenomenon as a desired outcome because we expect models to capture as many functions as possible. 

\begin{table}[!htb]
\centering
\caption{\textbf{Four model architectures with different combinations of feature pipelines.}}
\begin{tabularx}{\textwidth}{
	>{\raggedright\arraybackslash}X
	c
	c
	c
} 
\toprule
\toprule
& \multicolumn{3}{c}{Feature pipelines} \\
\cmidrule{2-4} 
Model  & Contact map & Correlation map & Persistence diagram 	\\
\midrule
Contact 		& \textbullet &  				& 				\\
ContactPers 	& \textbullet &  				& \textbullet 	\\
ContactCorr 	& \textbullet & \textbullet  	& 				\\
ContactCorrPers & \textbullet & \textbullet  	& \textbullet 	\\
\bottomrule
\bottomrule
\end{tabularx}
\label{table:models}
\end{table}

\begin{figure*}[htb!]
    \centering
    \includegraphics[width=\textwidth]{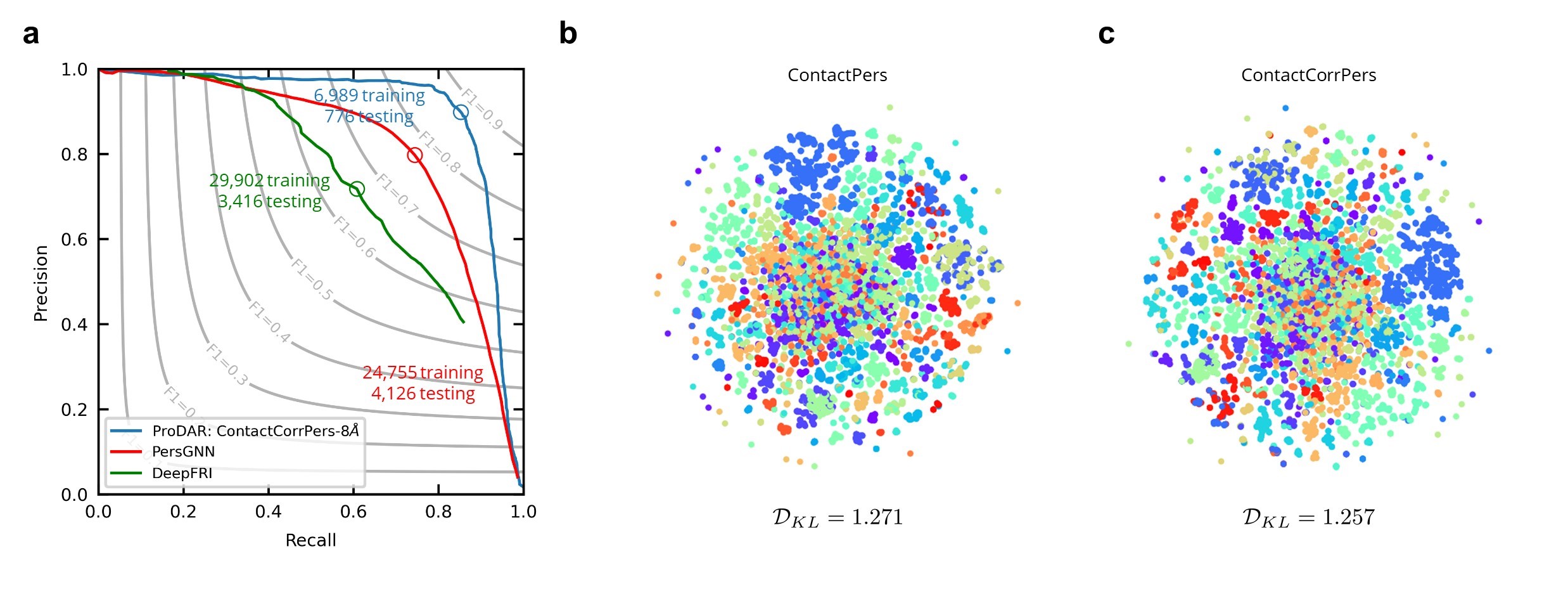}
    \caption{\textbf{Classification performance of ProDAR on MF-GO annotations. a} Precision-recall curves of DeepFRI, PersGNN, and ContactCorrPers network with GCN layers trained against $r_c = 8$ \AA\ dataset. 24,755 training PDBs and 4,126 testing PDBs  with $r_c = 8$ \AA\ are used by PersGNN. 29,902 training PDBs and 3,416 testing PDBs with $r_c = 10$ \AA\ are used by DeepFRI. Two-dimensional t-distributed stochastic neighbor embeddings (t-SNE) of protein representations predicted by \textbf{b} ContactPers and \textbf{c} ContactCorrPers networks. Each point represent a protein, whose color denotes the primary function predicted with the highest probability.}
    \label{fig:4}
\end{figure*}

Table \ref{table:scores} presents the classification performance of all models with different combinations of data pipelines. We evaluate the trained classifiers by leveraging the threshold at the final sigmoid activation. The area under the precision-recall curve (AUPR) and the maximum F1 score (F1-max) are calculated to evaluate the model performance. We find that GCNs in general outperform GraphSAGEs with maximum aggregators, despite a small margin of improvement. This agrees with the previous experimental and theoretical study showing that the mean aggregator essentially retains more discriminatory power than max aggregators since mean aggregator, although being more simplistic than the sum aggregator, learning the distribution of the node features such that it can be as powerful as the sum aggregator if the feature distribution is diverse or the recovery to the sum is possible\cite{xu2018powerful}. Our results also show that ContactPers and ContactCorr comparably enhance the representational capacity of the Contact model, with the ContactCorr model outperforming ContactPers by up to 0.005 in AUPR and 0.004 in F1-max score when GraphSAGE and the $r_c = 12$ \AA\ dataset are used. However, it is worth noting that by combining persistence and correlation together, the ContactCorrPers model has a notable performance gain that is greater than the direct sum of individual improvements by ContactPers and ContactCorr. The most robust classifier of the above models is ContactCorrPers with GCN convolutional layers for both the 8 \AA\ and 12 \AA\ datasets. Figure \ref{fig:4}a presents the precision-recall curves of ContactCorrPers nework, DeepFRI\cite{gligorijevic2021structure}, and PersGNN\cite{swenson2020persgnn}. Note that different datasets and contact cutoffs $r_c$ are used by three models. Figure \ref{fig:4}b and c illustrate the two-dimensional projections (t-distributed stochastic neighbor embedding, t-SNE) of final protein representation vectors by the ContactPers and ContactCorrPers networks. The Kullback–Leibler divergences between similarity probabilities at original and low-dimensional projections by the Contact and ContactCorr networks are 1.271 and 1.257, respectively, supporting that the ContactCorr network can better separate proteins in high dimensional space, and dynamics-informed protein representation can increase the discriminatory capacity of neural networks for protein function prediction. 

\begin{table}[!htb]
\centering
\caption{\textbf{Classification performance of ProDAR with different data pipelines on MF-GO annotations.} The best-performing models on  $r_c = 8, 12$ \AA\ datasets are highlighted in boldface.
}
\begin{tabularx}{\textwidth}{
	c
	>{\raggedright\arraybackslash}X 
	c
	c
} 
\toprule
\toprule
$r_c$ & Model  & AUPR & F1-max 	\\
\midrule
\multirow{9}{*}{8 \AA} 
	& Contact (GCN) 		& 0.898 	& 0.863 \\
 	& ContactPers (GCN) 	& 0.904		& 0.868	\\
 	& ContactCorr (GCN)		& 0.902  	& 0.867 \\
	& ContactCorrPers (GCN)	& \textbf{0.907}  	& \textbf{0.875} \\
\cmidrule{2-4}
	& Contact (GraphSAGE) 			& 0.876 	& 0.826 \\
 	& ContactPers (GraphSAGE) 		& 0.896		& 0.861	\\
 	& ContactCorr (GraphSAGE)		& 0.898  	& 0.861 \\
	& ContactCorrPers (GraphSAGE)	& 0.904  	& 0.871 \\
\midrule
\multirow{8}{*}{12 \AA} 
	& Contact (GCN) 		& 0.904 	& 0.879 \\
 	& ContactPers (GCN) 	& 0.910		& 0.878	\\
 	& ContactCorr (GCN)		& 0.910  	& 0.879 \\
	& ContactCorrPers (GCN)	& \textbf{0.916}  	& \textbf{0.895} \\
\cmidrule{2-4}
	& Contact (GraphSAGE) 			& 0.900 	& 0.871 \\
 	& ContactPers (GraphSAGE) 		& 0.905		& 0.873	\\
 	& ContactCorr (GraphSAGE)		& 0.910  	& 0.877 \\
	& ContactCorrPers (GraphSAGE)	& 0.915  	& 0.885 \\
\bottomrule
\bottomrule
\end{tabularx}
\label{table:scores}
\end{table}

\clearpage

\subsection*{DARs are critical functional residues with dynamic importance}

\begin{figure*}[htb!]
    \centering
    \includegraphics[width=\textwidth]{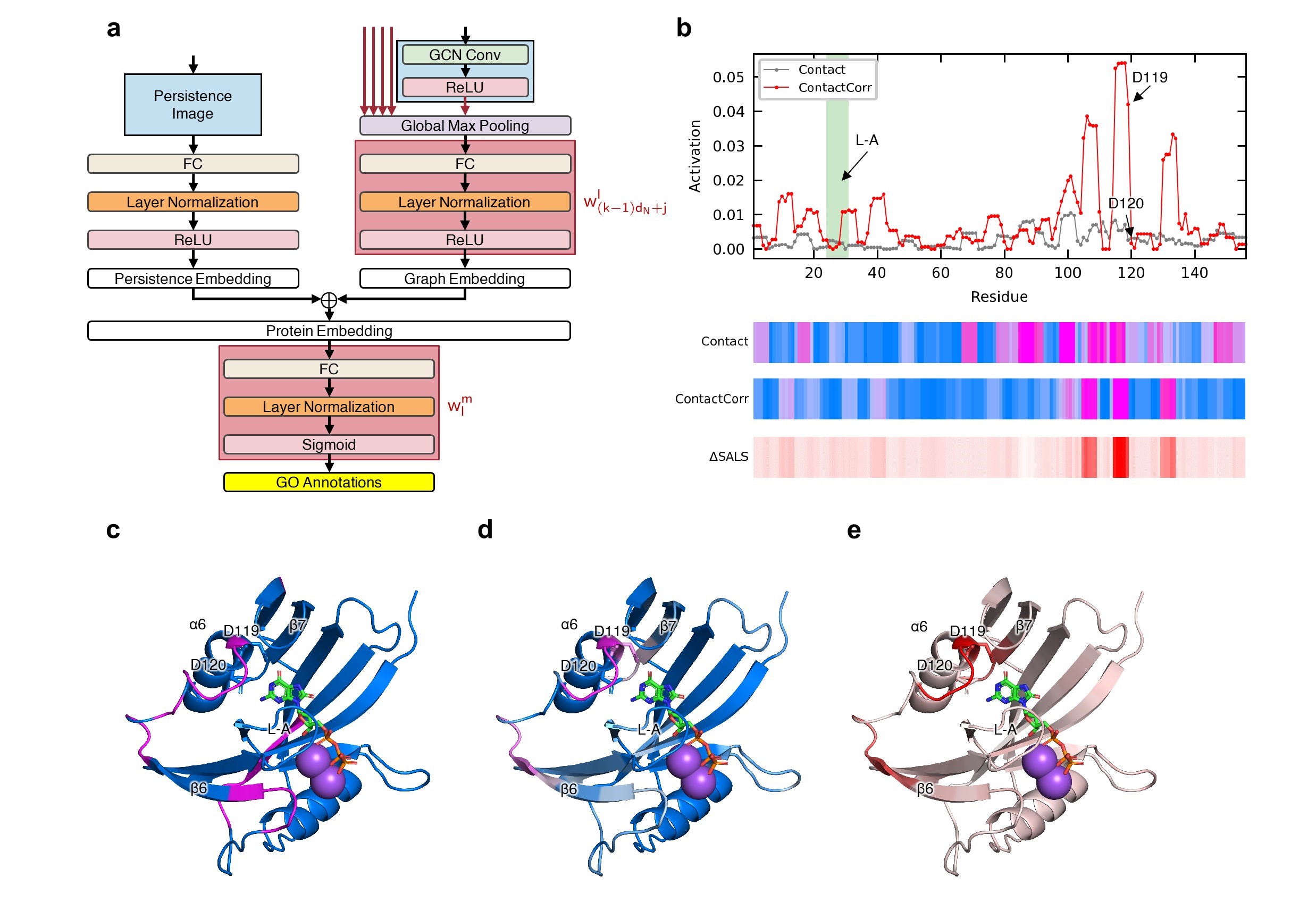}
    \caption{\textbf{DARs and functional residue identification. a} Schematic derivation of the importance weight $\mathbf{\alpha}^m_i$ on residue $i$ for function $m$. \textbf{b} Grad-CAMs of the Contact and ContactCorr networks for PDB 5GHI and the saliency difference ($\Delta\textrm{SALS}$) between the Grad-CAM obtained by the ContactCorr network and the Grad-CAM obtained by the Contact network. Green shaded area highlights loop L-A. \textbf{c-d} PDB structure annotations of activation maps predicted by \textbf{c} Contact and \textbf{d} ContactCorr networks. All residues are colored with a marine-magenta gradient according to the Grad-CAM profiles after min-max normalization, with more salient residues highlighted in magenta and less salient residues colored in marine blue. \textbf{e} PDB structure annotation of the saliency difference. All residues are colored with blue-white-red gradient. The residues colored in red and blue represent reinforcement and suppression, respectively, while the residues colored in white are comparably salient in both Grad-CAMs.}
    \label{fig:5}
\end{figure*}

Proteins are complex molecular machines whose functions are closely correlated to key functional residues. The experimental approaches usually include indirect methods, \textit{e.g.}, point mutation on specific residues, to identify the important collections of residues and their underlying functional mechanism. Examples abound in active sites of an enzyme, ligand-binding sites, ionic gating, protein-protein interaction (PPI), and mechanical conformation changes involving soft and stiff secondary structures\cite{gligorijevic2021structure}. On the computational side, many deep learning models have been applied to identify the functional residues in proteins\cite{torng2019high, forslund2008predicting}. To identify the functional residues with dynamic importance under graph description, we borrow the concept from class activation map (CAM), the method widely used in the field of computer vision to provide a visual explanation of convolutional neural networks\cite{zhou2016learning}. Specifically, we modify Grad-CAM\cite{selvaraju2017grad, gligorijevic2021structure}, which takes the derivatives of network outputs with respect to feature maps and either sums or averages the derivatives in spatial dimensions. The localized importance can be achieved by the sum of the feature maps weighted by the obtained derivatives (gradients). Unlike the original Grad-CAM averaging the derivatives via spatial pooling, in this work, we keep the vector form of derivatives and take the residuewise inner product of derivatives with residue-level embeddings across all graph convolutional layers (see Methods). This allows us to obtain residue-level activation for functional site identification while ensuring independence between two feature channdels. The importance weight of the feature map on residue $i$ for function $m$ ($\mathbf{\alpha}_i^m$) can thus be expressed as \begin{equation}
\boldsymbol{\alpha}_i^m = \sum_{l=1}^{d_G}\sum_{k=1}^{K}\sum_{j=1}^{d_N} w^m_l w^l_{(k-1)d_N + j} \mathbbm{1}\left(\underset{s}{\arg\max}\left\lbrace h_{sj}^k\right\rbrace = i \right)
\end{equation} where $w^l_{(k-1)d_N + j}$ is the equivalent weight from the GMP layer to the graph embedding $l$, and $w^m_l$ is the equivalent weight from graph embedding to the output layer, where we denote fully connected layer, layer normalization, and ReLU activation as a single operator (see the red blocks in Figure \ref{fig:5}a). $d_N$ and $d_G$ are node and graph embedding dimensions, respectively. $K$ is the total depth of graph convolutions. $\mathbbm{1}$ is the binary indicator function which equals 1 if the true condition is satisfied and 0 otherwise. For each PDB chain, we record the outputs activated by five graph convolutional layers (red arrow in Figure \ref{fig:5}a) and calculate the sum of the outputs weighted by the partial derivatives of GO annotations with respect to the feature channels in the convolutional layers. As the outputs feed forward through two fully connected layers with layer normalization and ReLU activation, the corresponding weights are hooked during calculations. 

Figure \ref{fig:5}b shows the Grad-CAMs of hMTH1 (PDB 5GHI) predicted by the Contact and ContactCorr networks at $r_c = 8$ \AA. While both networks activates residues around hydrogen bonding aspartate residues (Asp119 and Asp120) in their respective scales, only the ContactCorr network exactly activates Asp119 and mildly activates L-A, the loop surrounding the binding pocket (Figure \ref{fig:5}c-d). It is noticeable that ContactCorr network generally has larger activation profile than Contact network. To fairly compare two activation maps predicted by Contact and ContactCorr networks, an unbiased comparison without distorting the relative magnitudes of profiles is required. Direct subtraction is misleading under this circumstance because different models may have distinct scales of learned weights after training. Some models may only need small activation profiles for strong inferences. Min-max normalization fails as well since it inevitably rescales activation profiles from 0 to 1. Therefore, we compute the saliency maps predicted by Contact and ContactCorr networks first and subtract them afterward. As shown in Figure \ref{fig:5}b and e, the difference of the saliency maps ($\Delta\textrm{SALS}$, see Methods) emphasizes the loops and $\beta$-strand ends connected with $\alpha$-6 helix, including residues from Phe113 to Asp120, reflecting the dynamic significance of their roles in substrate binding\cite{waz2017structural}. The binding pocket changes conformation to recognize different substrates, as shown previously in Figure \ref{fig:1}b\cite{waz2017structural}. We speculate the flexibility of three activated regions enables the rotation of $\alpha$-6 helix to deform binding pocket for multiple specificity. Notably, the peaks of the activation map by the ContactCorr network are generally larger than those by the Contact network. This is attributed to the contribution of the correlation edges in the ContactCorr network. The correlation edges reinforce the activation of residues and provide more confident and stable functional inferences on graphs. These results show that with dynamic information in graph representation, ProDAR is able to identify the residues that are overlooked by the networks with mere input from the contact map.

\clearpage

\subsection*{ProDAR provides molecular insights into point mutations in proteins}

\begin{figure*}[htb!]
    \centering
    \includegraphics[width=\textwidth]{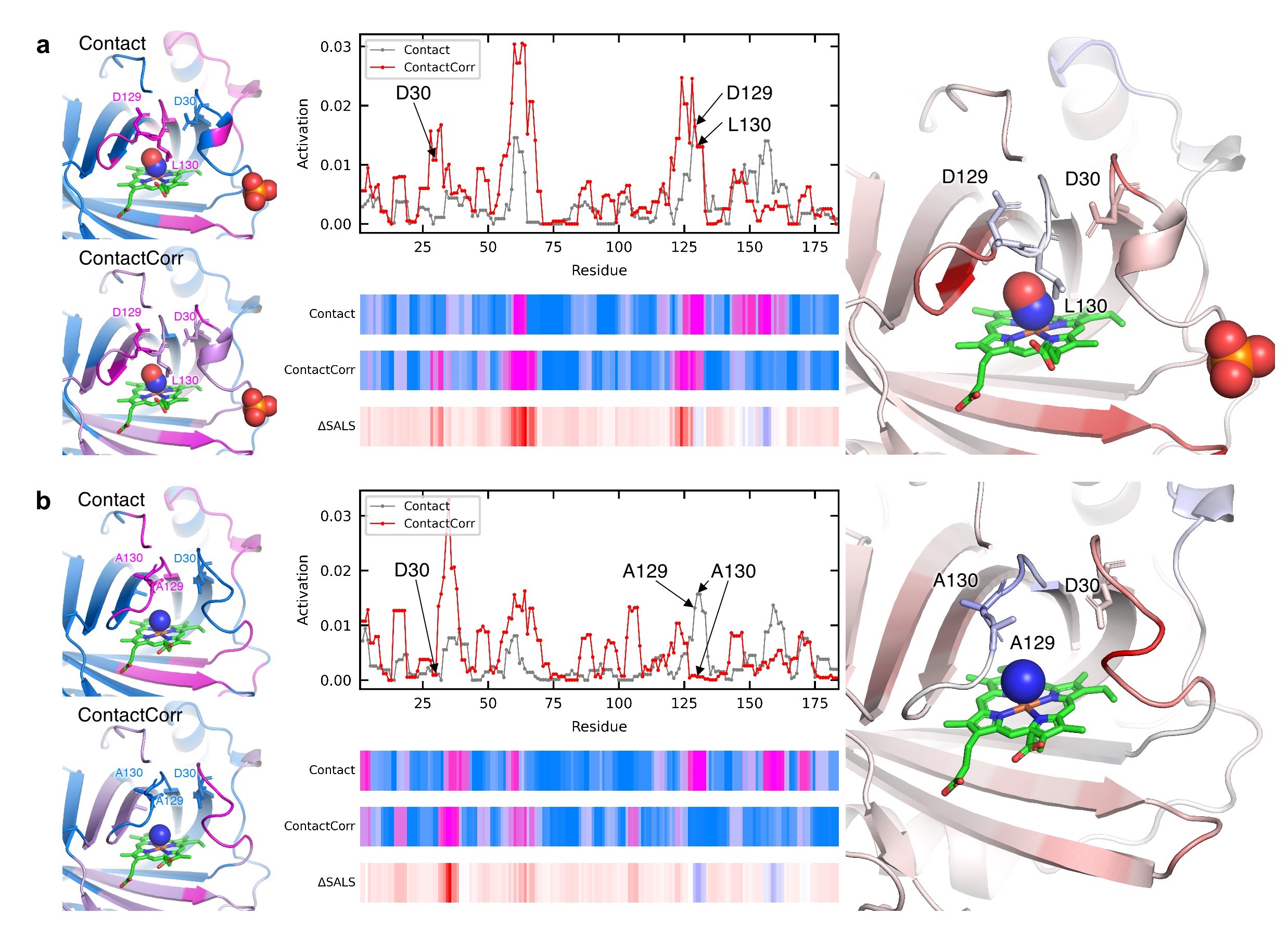}
    \caption{\textbf{Robustness and expressiveness of ProDAR to reflect point mutations in protein. a} Grad-CAMs and $\Delta\textsf{SALS}$ on T121V mutant of nitrophorin 4 (NP4) with nitric oxide, ferric heme, and phosphate ion (PDB 1SY1). \textbf{b} Grad-CAMs and $\Delta\textsf{SALS}$ on D129A/L130A mutant of NP4 with ferric heme and ammonium ion (PDB 1SY2). In Grad-CAMs, all residues are colored with marine-magenta gradient according to the Grad-CAM profiles after min-max normalization, with more salient residues highlighted in magenta and less salient residues colored in marine blue. In $\Delta\textsf{SALS}$, all residues are colored with blue-white-red gradient. The residues colored in red and blue represent reinforcement and suppression, respectively, while the residues colored in white are comparably salient in both Grad-CAMs. }
    \label{fig:6}
\end{figure*}

Figure \ref{fig:6} shows the Grad-CAMs of nitrophorin 4 (NP4)\cite{maes2004role}, a ferric heme protein that transports nitric oxide ($\textrm{NO}$) from blood-sucking insects to victims. The MF-GO terms of nitrophorin include oxidoreductase activity, metal ion binding, histamine binding, and nitric oxide binding. The A-B loop (residues 31-37) and G-H loop (residues 125-133) are indicated to be responsible for the pH dependence of $\textrm{NO}$ release\cite{maes2004role}. In the absence of $\text{NO}$ at high pH, the distal pocket in wild-type NP4 is open with poorly ordered A-B and G-H loops far from the heme. When $\textrm{NO}$ binds to wild-type NP4 at low pH, the distal packet collapses into a closed conformation, and the 130-131 peptide bond flips and forms a buried hydrogen bond between the Asp30 side chain and Leu130 carbonyl oxygen. Figure \ref{fig:6}a shows the activated residues of NP4 with a point mutant T121V (PDB 1SY1). Both the A-B and G-H loops are activated in the Grad-CAM predicted by the ContactCorr network, while only the G-H loop is activated by the Contact network. This proves that ContactCorr network can activate multiple residues within local chemical environment beyond contact cutoff distance. $\Delta\textrm{SALS}$ further indicates that residues around 30 and 125 are more salient in the ContactCorr activation. This agrees with the corresponding functional sites in NP4, where residues 30 and 125 are around A-B and G-H loops, respectively. Moreover, residues 120-124 are pertinent to hydrogen bonds with water in the distal pocket\cite{maes2004role}. In contrast, residues 150-159, the loop and $\alpha$-helix at the opposite outer side of heme binding sites, are suppressed. This implies that the ContactCorr network, unlike the Contact network, has a stronger denoising capability and does not overly activate trivial residues with more structural than functional purposes in protein.

In the structure of the D129A/L130A mutant of NP4, the 129-130 peptide bond is shifted by $\approx 2$ \AA\ , and Ala130 is rotated away from the heme, leading to an enlarged distal pocket for additional water molecules. This nullification of abilities for $\textrm{NO}$ binding and release is precisely reflected in the activation map by the ContactCorr network. By comparing the Grad-CAMs of the ContactCorr network in Figure \ref{fig:6}a and b, we note that Asp30, Ala129, and Ala130 are deactivated in the D129A/L130A structure. Moreover, the Grad-CAMs of the ContactCorr network are relatively more stable in the T121V and D129A/L130A structures. The two activation profiles exhibit similar activated sites. However, despite the ability to consistently identify some parts of functional residues (such as residues around 40 and 121), the Grad-CAMs of the Contact network show incoherent activation profiles for the other residues. In particular, it is unable to properly respond to the D129A/L130A mutations, leading to erroneous overactivation on peripheral loop (residues 168-175). Additionally, we note that N-terminus, which forms hydrogen bonds with Asp129 in the T121V structure and with Ala130 in the D129A/L130A structure, has larger activation values in the ContactCorr profiles than those in Contact profiles. This reveals that ProDAR has better sensitivity to the flexible terminals and the structural nuance associated with functional impacts.

\subsection*{Residue-level activation of ProDAR on the SARS-CoV-2 RBD}

\begin{figure*}[htb!]
    \centering
    \includegraphics[width=\textwidth]{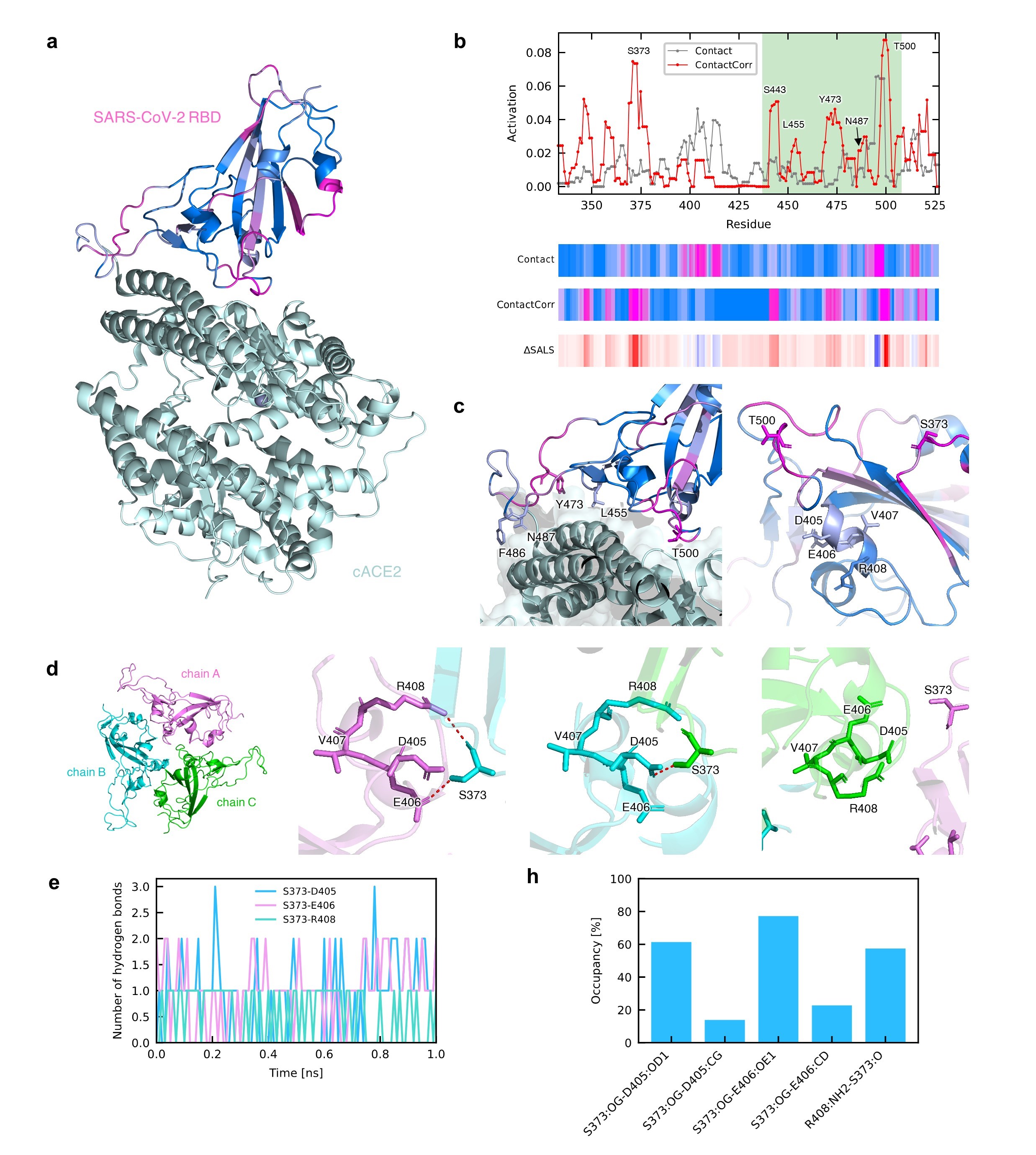}
    \caption{\textbf{DARs and hydrogen-bonding interactions in SARS-CoV-2 receptor-binding domain (RBD) a} Grad-CAM and $\Delta\textsf{SALS}$ of SARS-CoV-2 RBD (PDB 7C8D). \textbf{b} Residue activation predicted by ContactCorr network on SARS-CoV-2 RBD structure. Green shaded area highlights receptor-binding motif. \textbf{c} Complex structure of cat angiotensin
converting enzyme 2 (cACE2) bound to SARS-CoV-2 RBD. \textbf{d} Trimeric unit of SARS-CoV-2 spike glycoprotein (PDB 6VXX). Only SARS-CoV-2 RBD regions are visualized. The three panels on the right are hydrogen-bonding poses between A and B chains, B and C chains, and C and A chains, respectively. \textbf{e} Number of hydrogen bonds between SARS-CoV-2 RBDs over 1 ns MD simulation in the homotrimeric spike glycoprotein. Three donor-receptor pairs are counted simultaneously. \textbf{h} Hydrogen bond occupancy of each donor-receptor pair.}
    \label{fig:7}
\end{figure*}

At the forefront for the fight against the highly pathogenic severe acute respiratory syndrome coronavirus-2 (SARS-CoV-2), identifying residues of dynamic importance is urgently needed for structure-based vaccine design\cite{walls2020structure, lan2020structure}. Here, we apply ProDAR to the structure of the SARS-CoV-2 receptor binding domain (RBD) to determine whether our model is able to identify the important residues in the unseen and fragmented structure. Figure \ref{fig:7}a shows the complex of the SARS-CoV-2 RBD and cell receptor angiotensin converting enzyme 2 (ACE2) in cats\cite{wu2020broad}. Previous cryo-EM studies of the SARS-CoV spike protein and its interaction with ACE2 have shown that ACE2 receptor binding is a critical step for SARS-CoV to enter target cells\cite{yuan2017cryo}. However, the interaction between the SARS-CoV-2 RBD and ACE2 at the atomic level is not well understood. At the hydrophilic interface between the SARS-CoV-2 RBD and human ACE2 (hACE2), 13 hydrogen bonds and 2 salt bridges have been found previously\cite{lan2020structure}. Similar interactions have been found at the interface between the SARS-CoV-2 RBD and cat ACE2 (cACE2)\cite{wu2020broad}. In the complex of SARS-CoV-2 RBD and cACE2\cite{wu2020broad} (Figure \ref{fig:7}a), seven residues (Leu455, Phe456, Tyr473, Ser477, Phe486, Asn487, and Tyr489) on the $\beta 1'/\beta 2'$ loop of SARS-CoV-2 RBD have been found in contact with cACE2, including 2 hydrogen bonds (Ser477 with Gln18 of cACE2 and Asn487 with Tyr83 of cACE2\cite{wu2020broad}). As illustrated by Figure \ref{fig:7}, the receptor-binding motif of SARS-CoV-2 are highlighted by the green shaded area, where all seven residues identified in contacts with cACE2 are activated by the ContactCorr network. Other strongly activated residues, such as Ser443 and Thr500, are not identified by previous experiments in contact with cACE. In contrast, Contact network fails to activate four important binding sites Leu455, Phe456, Tyr473, and Ser477. Although neither the complex nor the individual atomic structure of SARS-CoV-2 RBD and cACE2 is in the training dataset and cACE2 is not present in the inference phase, the model yet exhibits strong predictive power and makes cogent inference on residues. 

In the Grad-CAM of the ContactCorr network, we find that there is a second largest peak around Ser373, which is not widely discussed in the literature. Molecular dynamics (MD) simulation is performed to confirm whether the residues around Ser373 are indeed crucial for protein functions. We simulate the full atomistic model of SARS-CoV-2 spike glycoprotein for 1 ns and find that Ser373 forms hydrogen bonds with other SARS-CoV-2 RBDs in the trimeric unit. In Figure \ref{fig:7}d, we show the hydrogen-bonding patterns between the SARS-CoV-2 RBDs in the entire homotrimeric spike glycoprotein\cite{walls2020structure}. Our results indicate that in the closed state, Ser373 tends to form hydrogen bonds with Asp405, Glu406, and Arg408 and stabilize the trimer of RBDs. At 0.77 ns, two and one hydrogen bonds form at the A-B and B-C interfaces, respectively, while no hydrogen bonds are found at the C-A interface. Figure \ref{fig:7}e shows the number of hydrogen bonds between SARS-CoV-2 RBDs. Three Ser373{\textendash}Asp405 hydrogen bonds form at the same time, while Arg408{\textendash}Ser373 pairs can form at most one hydrogen bond simultaneously. The occupancy analysis also shows that S373:OG-D406:OE1 is the most stable hydrogen bond, with occupancy close to 80\%, and that S373:OG-D405:OD1 and R408:NH2-S373:O have occupancy values of approximately 60\%, whereas S373:OG-D405:CG has only 13.86\% occupancy (Figure \ref{fig:7}h). The results again reveal that ProDAR can pinpoint the critical residues that have important functionality and molecular characteristics, even though the structure is not included in the training dataset. 

\clearpage
\section*{Discussion}

Through MD simulations, different conformations occurring at the nanosecond scale can be observed that, for instance, allow ion passing through the selectivity filters or encourage alternating conformations of transmembrane helices. On the other hand, PCA has also been increasingly used to extract the collective motions of macromolecules, whose thermal fluctuations conceal underlying contributions from highly cooperative movements that are closely related to their biological functions\cite{bahar2010normal}. NMA, a subgenre of PCA that provides a harmonic approximation of the energy landscape, has benefited this work and has enabled fast screening over an astronomical number of conformations for proteins.

The correlation map used in this work can be analogized to Pearson correlation coefficient. It has been indicated that Pearson correlation coefficient misses the orthogonal motions. When two highly correlated residues move perpendicularly, the correlation coefficient vanishes to zero and therefore becomes misleading even if two residues are strongly correlated\cite{lange2006generalized}. Generalized correlation coefficients (GCCs) has been conceived to resolve this issue by introducing Shannon mutual information\cite{kraskov2004estimating, lange2006generalized,melo2020generalized}. To calculate GCCs, multiple configurations in a time sequence are needed. The fluctuation vectors of normal modes lack the idea of time and can be scaled arbitrarily large in magnitude if temperature is not specified. Therefore, obtaining GCCs based on MD simulations is an important future direction for capturing more physically realistic correlations.

Although the correlation map derives from and combines modal frequencies and fluctuations in a single representation, it is possible to separately encode them. The modal frequencies can be separated as a feature vector for model input. Additionally, the mode fluctuations are a large set of vectors that can be manipulated by various mappings and operations. The correlation map is one possible pairwise reduction from the complete cross-correlation matrix. Other reduction approaches, such as the tensor invariant, infinity norm, and single value decomposition methods, can be employed. Helmholtz decomposition can also be used to compute the irrotational and solenoidal parts of mode fluctuations such that some rotational modes with significant physical meanings but modest inner product values can be successfully extracted.

It is believed that ANM unintentionally yields highly collective and large displacements concentrated in low frequency modes\cite{mendez2010torsional,dehouck2021large}. By performing eigenanalysis in torsion space, torsional network model (TNM) is able to delocalize the modes and avoid distorting covalent geometry of proteins\cite{mendez2010torsional}. Here we leave comparison between ANM, TNM, and other ENMs for future work. There are also other encoding methods from bioinformatics and deep learning, including the principal component score Vectors of Hydrophobic, Steric, and Electronic properties 8 (VHSE8)\cite{mei2005new}, blocks amino acid substitution matrices (BLOSUM62)\cite{eddy2004did}, and long short-term memory (LSTM)-based methods\cite{bepler2019learning, elabd2020amino}. This work nonetheless focuses on the augmented functional residue identification and the enhanced classification performance by incorporating dynamic information in deep neural networks. Other embedding methods of amino acids are temporarily out of the scope of this work but are indeed worthy of further experimentation for advanced inference problems, such as mutation detection and functional residue segmentation.

Different approaches for encoding dynamic information into graphs are of interest in future studies. The straightforward merging of contact and correlation edges can be replaced by more sophisticated graph architectures, \textit{e.g.}, dual graphs, directed graphs, and graphs with multiple types of edges. The values obtained from the correlation map can also be applied as the weights for edges, resembling the motifs of attention-based networks (for example, Graph ATtention network). While this work focuses on the utility of correlation edges and their implications for functional residue identification, we demonstrate that ProDAR has remarkable robustness against noise and has unprecedented interpretability for both regular and mutated PDB structures. Our model also shows strong generalizability to unseen PDB structures. Without \textit{a priori} knowledge of ligands and hydrogen bonds, ProDAR can still provide reliable inferences regarding PDB structures at the residue level. 

\section*{Conclusion}

In this work, we describe a method to encode the dynamic information of proteins in graph representation. It is the first time to incorporate the NMA of proteins into a deep learning framework for functional prediction. In addition to the contact map commonly used by the previous graph-based deep learning models\cite{swenson2020persgnn, gligorijevic2021structure, sanyal2020proteingcn}, we compute the correlation map and add edges between the residue pairs that are highly correlated in the first 20 modes. These correlation edges connect the dynamically correlated residues distant in space and allow message to pass over long spatial distances. The model therefore alters the graph inference regime and better captures the dynamic information intrinsically residing in the native conformation of proteins. Such dynamics-informed graph representation exhibits consistent performance gains in protein function classification and increases the discriminatory power of neural networks to differentiate the structural nuance in proteins. ProDAR, with input of dynamic information from NMA, identifies DARs that are both structurally and dynamically influential on protein function. 

To summarize, we close the information gap between protein structure and function by introducing dynamic analysis in the graph construction process. Our method bridging structure and function with dynamics provides the missing piece of the deep learning framework for protein function prediction and sketches a comprehensive picture from sequence to function, with great potential to advance protein science in mutation detection and function engineering through rapid screening over large sequential, structural, and dynamical spaces.

\clearpage
\section*{Methods}
\subsection*{Resource availability}
\subsubsection*{Lead contact}
Further information and requests for resources and reagents should be directed to and will be fulfilled by the lead contact, Shu-Wei Chang (\href{mailto:changsw@ntu.edu.tw}{changsw@ntu.edu.tw}).

\subsubsection*{Materials availability}
This study did not generate new unique materials.

\subsubsection*{Data and code availability}
The code implementation associated with this paper is publicly available on GitHub
(\href{https://github.com/chiang-yuan/ProDAR}{https://github.com/chiang-yuan/ProDAR})

\subsection*{Normal mode analysis (NMA)}

Given a surrogate physical system for a protein represented by $N$ interaction particles ($\textrm{C}^\alpha$ atoms), the potential energy can be expressed as the Taylor expansions near the native conformation $\mathbf{q}^o$. By omitting higher order term to the second order, the required energy to migrate from $\mathbf{q}^o$ to $\mathbf{q}$ becomes the sum of the pairwise potentials \begin{equation}
\label{eq:eng}
\begin{aligned}
V(\mathbf{q}) - V(\mathbf{q}^o) &= \sum_{i=1}^{3N}\sum_{j=1}^{3N}\left.\frac{\partial^2 V}{\partial q_i \partial q_j}\right|_{\mathbf{q}^o}(q_i - q_i^o)(q_j - q_j^o) \\ &= \sum_{i=1}^{3N}\sum_{j=1}^{3N}H_{ij}(q_i - q_i^o)(q_j - q_j^o) = \frac{1}{2} \Delta\mathbf{q}^\intercal\mathbf{H}\Delta\mathbf{q}
\end{aligned}
\end{equation} where $\mathbf{H}$ is the $3N\times 3N$ Hessian matrix of the second derivatives of the potential with respect to particle coordinates at equilibrium \begin{equation}
H_{ij} = \left.\frac{\partial^2 V}{\partial q_i\partial q_j}\right|_{\mathbf{q}^o}
\end{equation}. By considering only the internal interactions under the classical regime, the equation of motion of particles can be written as \begin{equation}
\label{eq:eom}
\mathbf{M}\Delta\ddot{\mathbf{q}} + \mathbf{H}\Delta\mathbf{q} = \mathbf{0}
\end{equation}, where $\mathbf{M}$ is the diagonal matrix of particle masses. The solution to equation \eqref{eq:eom} is a $3N$-dimensional harmonic oscillator, which can be further generalized as \begin{equation}
\label{eq:sol}
\Delta\mathbf{q}^{(k)} = \mathbf{u}^{(k)} = \mathbf{a}^{(k)}\exp\left(-i\omega_k t\right)
\end{equation} where $\mathbf{a}^{(k)}$ is a complex vector containing the amplitude and phase factor, and $\omega_k$ is the frequency of the $k$-th solution. Substituting the solution (equation \eqref{eq:sol}) into equation \eqref{eq:eom} gives the generalized eigenvalue equation \begin{equation}
\mathbf{H}\mathbf{u}^{(k)} = \omega_k^2 \mathbf{M}\mathbf{u}^{(k)}
\end{equation}. By stacking a complete set of solution vectors into a single matrix $\mathbf{U}$ and arranging the corresponding squared frequencies as a diagonal matrix $\boldsymbol{\Lambda}$, the equation \eqref{eq:sol} can be rewritten as \begin{equation}
\label{eq:eigen}
\mathbf{H}\mathbf{U} = \mathbf{M}\mathbf{U}\boldsymbol{\Lambda}
\end{equation}. The equation \eqref{eq:eigen} can be solved by transforming it into a standard eigenvalue equation through mass-weighed transformations: \begin{align}
\tilde{\mathbf{U}} &= \mathbf{M}^{\frac{1}{2}}\mathbf{U} \\
\tilde{\mathbf{H}} &= \mathbf{M}^{-\frac{1}{2}}\mathbf{H}\mathbf{M}^{-\frac{1}{2}}
\end{align}. Thus the massed-weighted eigenvalue equation becomes \begin{equation}
\label{eq:masseigen}
\tilde{\mathbf{H}}\tilde{\mathbf{U}} = \tilde{\mathbf{U}}\boldsymbol{\Lambda}
\end{equation}. The equation \eqref{eq:masseigen} has $3N-6$ non-zero eigenvalues, reflecting six degenerate eigenstates of three translational and three rotational degrees of freedom. The mass-weighed Hessian $\tilde{\mathbf{H}}$ remains real, symmetric, and positive semi-definite as the original Hessian $\mathbf{H}$ by construction. Its eigenvectors $\tilde{\mathbf{u}}^{(k)}$ (column vectors of $\tilde{\mathbf{U}}$, $k = 1, 2, \cdots, 3N$ ) form an orthonormal basis and construct normal modes of the system in the mass-weighed coordinates. The original normal modes in Cartesian coordinates can hence be obtained by $\mathbf{U} = \mathbf{M}^{-\frac{1}{2}}\tilde{\mathbf{U}}$. Since  $\tilde{\mathbf{U}}$ is a unitary matrix, the orthogonality is again satisfied by \begin{equation}
\mathbf{I} = \tilde{\mathbf{U}}^\intercal\tilde{\mathbf{U}} = \mathbf{U}^\intercal\left(\mathbf{M}^{\frac{1}{2}}\right)^\intercal \mathbf{M}^{\frac{1}{2}}\mathbf{U} = \mathbf{U}^\intercal\mathbf{M} \mathbf{U}
\end{equation}. The energy change associated with a given mode $k$ is then proportional to the square of its mode frequency $\omega_k$, as seen in equation \eqref{eq:eng}: \begin{equation}
V\left(\mathbf{u}^{(k)} = \Delta\mathbf{q}^{(k)}\right) = \frac{1}{2}\left(\mathbf{u}^{(k)}\right)^\intercal\mathbf{H}\mathbf{u}^{(k)} = \frac{\omega_k^2}{2}
\end{equation} Fluctuations along the high-frequency modes are therefore energetically more expensive than those along the low-frequency modes. According to the equipartition theorem, the vibrational energy is prone to be equally partitioned among all the modes, such that the vibrational amplitudes are scaled by $1/\omega_k^2$. 

NMA is based on the assumption of symmetric and positive semi-definite Hessian matrix. Strictly speaking, energy minimization needs to be carried out before performing NMA on protein crystal structure to ensure the local energy minimum. However, energy minimization is computationally intensive. An alternative method is adopting elastic network model (ENM) that accept the initial structure (crystal structure from PDB) to be the conformation at local energy minimum. 

\subsection*{Anisotropic network model (ANM)}

Anisotropic network model (ANM) is the most broadly used ENM\cite{tirion1996large, atilgan2001anisotropy, eyal2006anisotropic, bahar2010normal}. We first extracted $\text{C}^\alpha$ atoms of amino acids as nodes and connected them with springs if the separating distance lies within the prescribed cutoff distance $r_c$. The potential energy of ANM is given by the sum of pairwise harmonic potentials: \begin{equation}
V(\mathbf{q}) = \frac{1}{2}\sum_{\alpha,\beta} \gamma_{\alpha\beta} \left(r_{\alpha\beta} - r_{\alpha\beta}^o\right)^2 
\end{equation} where $\gamma_{\alpha\beta}$ is the spring force constant between $i, j$ pair of atoms; $r_{\alpha\beta} = \|\mathbf{q}_\beta - \mathbf{q}_\alpha\|$ and $r_{\alpha\beta}^o = \|\mathbf{q}_\beta^o - \mathbf{q}_\alpha^o\|$ are the instantaneous distance and the equilibrium distance between two atoms, respectively. Here we use greek letters as indices to avoid confusion with the indices of individual components used in equation \eqref{eq:eng}. The components in the Hessian of ANM $\mathbf{H}$ can be obtained from the second derivatives with respect to the components of $\alpha$ and $\beta$ atoms: \begin{equation}
\frac{\partial^2 V}{\partial x_\alpha \partial y_\beta} = -\frac{\gamma_{\alpha\beta}}{r_{\alpha\beta}^2} \left(x_\beta - x_\alpha\right)\left(y_\beta - y_\alpha\right) \frac{r_{\alpha\beta}^o}{r_{\alpha\beta}}
\label{eq:vxy}
\end{equation} where $x, y$ are free indices interchangeable with respect to the three dimensions of coordinates. Evaluating equation \eqref{eq:vxy}  at equilibrium conformation $\mathbf{q}_\alpha^o, \mathbf{q}_\beta^o$ yields the off-diagonal Hessian submatrices \begin{equation}
\mathbf{H}_{\alpha\beta} =  -\frac{\gamma_{\alpha\beta}}{{r_{\alpha\beta}^o}^2} \begin{bmatrix}
\left(x_\beta^o - x_\alpha^o\right)^2 &  \left(x_\beta^o - x_\alpha^o\right)\left(y_\beta^o - y_\alpha^o\right) &  \left(x_\beta^o - x_\alpha^o\right)\left(z_\beta^o - z_\alpha^o\right) \\
\left(x_\beta^o - x_\alpha^o\right)\left(y_\beta^o - y_\alpha^o\right) &  \left(y_\beta^o - y_\alpha^o\right)^2 &  \left(y_\beta^o - y_\alpha^o\right)\left(z_\beta^o - z_\alpha^o\right) \\
\left(x_\beta^o - x_\alpha^o\right)\left(z_\beta^o - z_\alpha^o\right) &  \left(y_\beta^o - y_\alpha^o\right)\left(z_\beta^o - z_\alpha^o\right) &  \left(z_\beta^o - z_\alpha^o\right)^2 \\
\end{bmatrix}
\end{equation} and the diagonal Hessian submatrices \begin{equation}
\mathbf{H}_{\alpha\alpha} = - \sum_{\beta\in\mathcal{N}(\alpha)\setminus \alpha} \mathbf{H}_{\beta\alpha}
\end{equation}. The spring force constants $\gamma_{\alpha\beta}$ are set as 1 in this work.

To elucidate the dynamic behavior of proteins about the equilibrium conformation and find out the underlying cooperative motions, it is of interest to know the cross-correlations between residue fluctuations given by the mode shapes, \textit{i.e.} the eigenvectors of the Hessian, within and across different normal modes. To this end, we can calcualte the ensemble average of cross-correlations between certain components of eigenvectors \begin{equation}
\left\langle \Delta q_i \Delta q_j\right\rangle = \frac{1}{Z} \int d^{3N}q e^{-\frac{\Delta\mathbf{q}^\intercal\mathbf{H}\Delta\mathbf{q}}{2k_BT}} \Delta q_i \Delta q_j = k_BT\left[\mathbf{H}^{-1}\right]_{ij}
\end{equation} where $Z$ is the configurational integral \begin{equation}
Z = \int d^{3N}q e^{-\frac{\Delta\mathbf{q}^\intercal\mathbf{H}\Delta\mathbf{q}}{2k_BT}} = \left(2\pi k_B T\right)^\frac{3N}{2} \left[\det\left(\mathbf{H}^{-1}\right)\right]^\frac{1}{2}
\end{equation}, and $\left[\mathbf{H}^{-1}\right]_{ij}$ is the $i,j$-th element in the inverse of Hessian. The integration performs over the entire configurational space following Boltzmann distribution and therefore favors low energy modes with high probability over high energy modes. By equation \eqref{eq:masseigen}, the inverse of Hessian can be expressed as \begin{align*}
\tilde{\mathbf{H}} &= \tilde{\mathbf{U}}\boldsymbol{\Lambda}\tilde{\mathbf{U}}^\intercal \\ 
\tilde{\mathbf{H}}^{-1} &= \left(\tilde{\mathbf{U}}\boldsymbol{\Lambda}\tilde{\mathbf{U}}^\intercal \right)^{-1} = \left(\tilde{\mathbf{U}}^\intercal\right)^{-1} \boldsymbol{\Lambda}^{-1} \tilde{\mathbf{U}}^{-1} =  \tilde{\mathbf{U}} \boldsymbol{\Lambda}^{-1} \tilde{\mathbf{U}}^\intercal
\end{align*}. Since there are six degenerate modes with zero eigenvalues, the inverse of Hessian is replaced by the \textit{pseudoinverse}, which takes the sum of non-degenerate modes \begin{equation}
\tilde{\mathbf{H}}^{-1} = \sum_{k=1}^{3N-6} \frac{\tilde{\mathbf{u}}^{(k)}\left(\tilde{\mathbf{u}}^{(k)}\right)^\intercal}{\omega_k^2}
\end{equation}. Therewith we have the cross-correlation matrix that is the product of inverse eigenvalues and eigenvectors by transforming it into the Cartesian coordinates \begin{equation}
\mathbf{C} = k_B T {\mathbf{H}}^{-1} = k_B T \sum_{k=1}^{3N-6} \frac{{\mathbf{u}}^{(k)}\left({\mathbf{u}}^{(k)}\right)^\intercal}{\omega_k^2}
\end{equation}. It should be noted that $\mathbf{C}$ has size of $3N \times 3N$ and gives the cross-correlation between the individual components of atom coordinates. We therefore reduce it into $N \times N$ matrix $\mathbf{c}$ by taking the trace of the $3 \times 3$ submatrices of $\mathbf{C}$, which is equivalent to the sum of the inner products of eigenvectors between $i, j$ residues: \begin{equation}
\label{eq:corr}
c_{ij} = k_B T \textrm{tr}\left(\sum_{k=1}^{3N-6} \frac{{\mathbf{u}_i}^{(k)}\left({\mathbf{u}_j}^{(k)}\right)^\intercal}{\omega_k^2}\right) = k_B T \sum_{k=1}^{3N-6} \frac{\left(\mathbf{u}_i^{(k)}\right)^\intercal\mathbf{u}_j^{(k)}}{\omega_k^2} \approx k_B T \sum_{k=1}^{20} \frac{\left(\mathbf{u}_i^{(k)}\right)^\intercal\mathbf{u}_j^{(k)}}{\omega_k^2}
\end{equation}, where we remove degenerate modes and follow previous PCA experiments\cite{david2014principal} to truncate all available modes to the first 20 modes for subsequent calculation. To verify 20 modes are sufficient, we evaluate the individual contribution of each mode by calculating the proportion of the inverse eigenvalue to all modes: \begin{equation}
\label{eq:prop}
\frac{1/\lambda_i}{\sum_{k=1}^{3N-6}1/\lambda_k} = \frac{1/\omega_i^2}{\sum_{k=1}^{3N-6}1/\omega_k^2}
\end{equation}, where $1/\omega_i^2$ is proportional to the vibrantional amplitude of the mode. 

Note that all elements of cross-correlation matrix $\mathbf{c}$ are linearly proportional to $k_BT$ and thus can be canceled out after normalization with respect to each residues. We therefore reach our final normalized cross-correlation matrix (correlation map) \begin{equation}
\bar{c}_{ij} = \frac{c_{ij}}{\sqrt{c_{ii}c_{jj}}}
\end{equation} such that all of the diagonal elements are equal to 1 and off-diagonal elements represent the normalized correlation between residues. 

\subsection*{Persistent homology (PH)}

Persistent homology (PH) is an algebraic topology method for measuring topological features of shapes, data, and functions\cite{edelsbrunner2008persistent, edelsbrunner2010computational}. It provides a stable tool for a myriad of applications, including biomolecules classification, geometric modeling, and network analysis.

Consider the union of identical $\epsilon$-balls around $\text{C}^\alpha$ atoms. The simplicial complex (points, lines, triangles, and terahdra) of $\text{C}^\alpha$ atoms start to emerge and then disappear (birth and death of $k$-simplices) as the radii of the balls increase. The nested sequence of this simplicial complex is the \textit{filtration} filtered by the radius of the ball. Each birth-death pair is then transformed into the \textit{persistence diagram} (PD) by mapping to birth-persistence coordinate. The vertical distance between diagonal line and point is exactly the persistence of the topological feature. We used \textit{alpha complex} to compute the one- and two-dimensional persistent homology group $H_1, H_2$. For a given $\alpha$, the alpha complex includes all the simplices in Delaunay triangulation which have an empty circumscribing ball with squared radius equal or smaller than $\alpha$. The filtration of alpha complex therefore allows us to extract the topological features of loops and cavities in protein structures.

\subsection*{Persistence images (PI)}

Let $D$ be a PD in birth-death coordinates and $T: \mathbb{R}^2 \rightarrow \mathbb{R}^2$ be the linear transformation $T(x_1,x_2) = (x_1, x_2-x_1)$ from birth-death coordinates to birth-persistence coordinates. At each transformed point $\mathbf{u}=(u_1, u_2)  = (x_1, x_2-x_1)\in T(D)$ we placed a Gaussian kernel function \begin{equation}
\phi_{\mathbf{u}, \sigma^2}(\mathbf{z}) = \frac{1}{2\pi\sigma^2}e^{-\frac{(z_1 - u_1)^2 + (z_2 - u_2)^2}{2\sigma^2}}
\label{eq:gaussian}
\end{equation} where $\sigma^2$ is the variance of the gaussian kernel. With an appropriate weighting function $w: \mathbb{R}^2 \rightarrow \mathbb{R}$, we transform PD into the \textit{persistence surface} \begin{equation}
\rho_D(\mathbf{z}) = \sum_{\mathbf{u} \in T\left(D\right)} w(\mathbf{u}) \phi_{\mathbf{u}, \sigma^2}(\mathbf{z}) 
\end{equation}. We designate $\sigma^2 = 1$ and $w$ as the square of persistence \begin{equation}
w(\mathbf{u}) = u_2^2 = \left(x_2 - x_1\right)^2
\end{equation}. The function intensifies the feature with larger persistence and satisfies stability requirements being zero along $z_1$, continuous, and piecewise differentiable\cite{edelsbrunner2008persistent}. By taking the pixel-wise integral of the discretized persistence surface, we obtain the \textit{persistence image} (PI) consisting of a vector of integral value for each pixel $p$ \begin{equation}
I(\rho_D)_p = \iint_p \rho_D dxdy
\end{equation}. Throughout this work, we transform the identical PD domains within $[0, 50] \times [0, 50]$ into PIs of size $25 \times 25$, namely vectors of 625 dimensions. The choices of image resolution, kernel probability distribution, as well as weighting functions may vary for different tasks at hands and should be left as an open question to select the optimal ones. However, previous experimental studies have indicated that the classification accuracy in machine learning framework is robust against the choice of image resolution\cite{zeppelzauer2016topological, edelsbrunner2008persistent} and has low sensitivity to the variance of Gaussian kernel (equation \eqref{eq:gaussian}). The implementation of persistent homology and persistence image representation were carried out using GUDHI library with CGAL as the backend for alpha complex calculation\cite{maria2014gudhi, cgal:dy-as3-21b}. 

\subsection*{Datasets and training configurations}

Our dataset consists of 7,765 proteins and 155 Gene Ontology (GO) annotations retrieved from Protein Data Bank (PDB)\cite{berman2003announcing, mir2018pdbe, burley2021rcsb, kinjo2016protein, kinjo2018new} and Structure Integration with Function, Taxomony, and Sequence (SIFTS) database\cite{velankar2012sifts, mir2018pdbe, dana2019sifts}. The 3D atomic structures of proteins were collected based on the size and resolution. We filtered out proteins larger than 5,000 deposited atoms excluding water molecules, and selected the proteins starting from the highest structure resolution. Among approximate 110,000 legitimate protein candidates, the first 10,000 proteins sorted by resolution were first selected, with 7,765 of them (which have valid PDB, contact maps, and SIFTS MF-GO annotations) retrieved through ProDy package\cite{bakan2011prody, cock2009biopython}. ProDy is a protein structural and dynamical analysis tool streaming multiple resources of PDBs affiliated with World Wide Protein Data Bank (wwPDB)\cite{berman2003announcing, mir2018pdbe, burley2021rcsb, kinjo2016protein, kinjo2018new}. Among 7,765 proteins, 6,989 of them are used for training and validation, and the remaining 776 proteins are used for testing. Protein graphs are processed using NetworkX package\cite{hagberg2008exploring}. 3D protein structures are visualized by PyMOL\cite{PyMOL} and VMD\cite{humphrey1996vmd} in this work. We trained our model to predict the MF-GO annotations that have enough training samples. The MF-GO terms that annotate less than 25 PDB entries are excluded from our dataset. We summarize the 20 largest CATH families and 25 largest MF-GO classes in Figure \ref{fig:cath} and \ref{fig:topgoterms}, respectively. 

To optimize the multi-label classification prediction and take the imbalanced samples into account, all models were trained to minimize the weighted cross-entropy loss function that induces models to act if the balanced samples are given: \begin{equation}
\mathcal{L} = -\frac{1}{NL}\sum_{i=1}^N\sum_{j=1}^L w_j y_{ij} \log\left(\hat{y}_{ij}\right) + (1 - y_{ij}) \log\left(1 - \hat{y}_{ij}\right)
\end{equation} where $N$ is the total number of training samples, $L$ is the number of MF-GO classes, $y_{ij}$ is the true binary indicator for protein $i$ to have MF-GO function $j$, and $\hat{y}_{ij}$ is the predicted probability that protein $i$ is annotated with MF-GO function $j$; $w_j$ is the added weight for the positive class $j$, which is assigned as the ratio between negative and positive samples \begin{equation}
w_j = \frac{n_j^-}{n_j^+}
\end{equation}

At inference stage, the protein $i$ is said to have function $j$ annotated with associated GO terms if the predicted probability $\hat{y}_{ij} \geq 0.5$. True positive rate (TPR), true negative rate (TNR), false positive rate (FPR), and false negative rate (FNR) are calculated by varying the threshold value from 0 to 1 to obtain the precision-recall curve. 

We randomly separated our dataset into train and test set by 90\% and 10\%. We used ADAM optimizer\cite{kingma2014adam} with learning rate $\text{LR} = 5\times 10^{-5}$ and exponential decay rate $\beta_1 = 0.9, \beta_2 = 0.999$. To avoid overfitting, we employed weight decay $1 \times 10^{-5}$ and further applied dropout regularization $p=0.1$ and layer normalization after every linear layer\cite{loshchilov2017decoupled,ba2016layer}. Each model is trained for 300 epochs with batch size of 64 on NVIDIA V100 16/32G GPU. The model training is implemented using PyTorch and PyTorch Geometric library\cite{fey2019fast}.

To ensure our models do not overfit to the training data, we have further experimented five-fold cross-validation on ContactCorr network by splitting all data into training, validation, and testing sets at a ratio of 72\%, 18\%, 10\%. All five models have comparable AUPR and F1-max metric scores (see Figure \ref{fig:plot-kfold}).

\subsection*{Residue-level activation map and saliency difference $\Delta\textrm{SALS}$}

To locate the residues contributing to the predicted function labels, we devise a method inspired by Gradient-weighted Class Activation Map (Grad-CAM)\cite{selvaraju2017grad, gligorijevic2021structure} to find the residues with highest contribution at inference stage. Motivated by its recent success in image classification and residue-level annotations for proteins, we used Grad-CAM to identify the residues that are important for function prediction. The method produces visual explanation based on neural networks and highlights the features that contribute more to the activation in the downstream neurons.

We use PyTorch hook to extract the residue embeddings after each graph convolutional layer $\mathbf{H}^{(k)} \in \mathbb{R}^{V\times D_k}$, where $V$ is the number of residues and $D_k$ is the dimension of residue embedding at $k$-th layer. To compute the contribution of each residue $i$ to the prediction of function $j$, we compute derivatives of the model output layer $\hat{y}_j$ with respect to residue embeddings $\mathbf{h}_{i}^{(k)} \in \mathbb{R}^{D_k}$: \begin{equation}
\boldsymbol{\alpha}_i^j = \frac{\partial \hat{y}_j}{\partial \mathbf{h}_i}
\end{equation} where $\boldsymbol{\alpha}^j$ measures the importance weight of each feature in residue embeddings for predicting function $j$. Instead of summing the derivatives from individual residue for measuring the importance of a specific feature map $\mathbf{f}_c \in \mathbb{R}^{V}$ (\textit{i.e.} $c$ channel of the residue embedding along residue space, $\mathbf{H}^{(k)} = \left[\mathbf{f}_1, \cdots, \mathbf{f}_{D_k}\right] = \left[\mathbf{h}_1, \cdots, \mathbf{h}_{V}\right]^\intercal$) as implemented by Gligorijevi\'{c} \textit{et al.}\cite{gligorijevic2021structure}, we maintained the vector form of feature map and took its inner product with the importance vector for each residue to obtain the funtion-specific heatmap in the residue space: \begin{equation}
\text{CAM}^j_i = \text{ReLU}\left(\boldsymbol{\alpha}_i^j \cdot \mathbf{h}_i \right) = \text{ReLU}\left(\sum_{c=1}^{D_k} \alpha_{ic}^j h_{ic} \right)
\end{equation} where $\text{ReLU}$ function ensures that only positive contributions to the function prediction are preserved. 

The difference of saliency map is given by: \begin{align}
\Delta\textsc{SALS} &= \textsc{SALS}^\text{ContactCorr} - \textsc{SALS}^\text{Contact} \\
\textsc{SALS}_i &= \sum_{j = 1}^{V} | I_j - I_i |
\end{align} where $I_j$ is the intensity of activation signal at residue $j, \forall j \in \mathcal{V}$. The saliency map accentuates the region that is noticeable among its neighborhood, providing a self-referential measure of feature importance such that the comparison between the Grad-CAM from ContactCorr network and the vanilla one from Contact network becomes meaningful. The difference of saliency map $\Delta\textrm{SALS}$ therefore identifies the residues that are confidently activated or suppressed by the ContactCorr network in comparison with Contact network. 

\section*{Acknowledgements}
The authors appreciate the financial support from the Ministry of Science and Technology, Taiwan [109-2224-E-007-003, 110-2636-E-002-013]. We thank National Center for High-performance Computing (NCHC) for providing computational and storage resources.

\section*{Author Contributions}
Y.C. and S.W.C. conceived the research ideas; Y.C. designed, developed, and trained the deep learning model; W.H.H. performed the molecular dynamics simulation; Y.C. analyzed the simulation results; Y.C. wrote the paper with input and advice from S.W.C.

\section*{Declaration of Interests}
The authors declare no competing interests.

\bibliography{dar}

\clearpage

\renewcommand{\appendixname}{Supplementary Information}
\appendix
\renewcommand{\thesection}{Supplementary Information \arabic{section}}
\renewcommand{\thefigure}{S\arabic{figure}}
\setcounter{figure}{0}
\renewcommand{\thetable}{S\arabic{table}}
\setcounter{table}{0}
\pagenumbering{gobble}

\section*{Supplementary Information}

\vfill
\begin{figure*}[htb!]
    \centering
    \includegraphics[width=\textwidth]{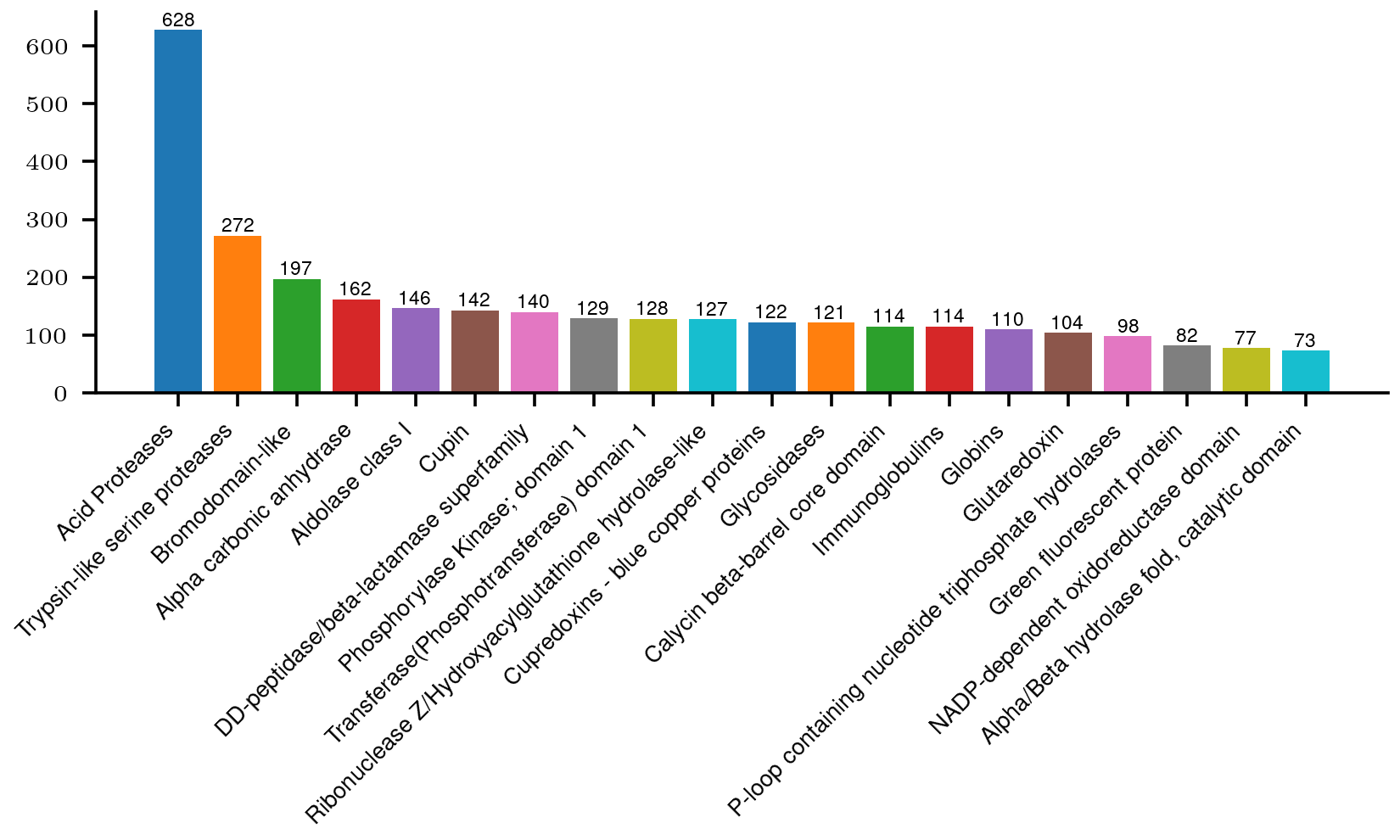}
    \caption{Summary of 20 largest CATH families in our dataset.}
    \label{fig:cath}
\end{figure*}
\vfill

\begin{figure*}[htb!]
    \centering
    \includegraphics[width=0.7\textwidth]{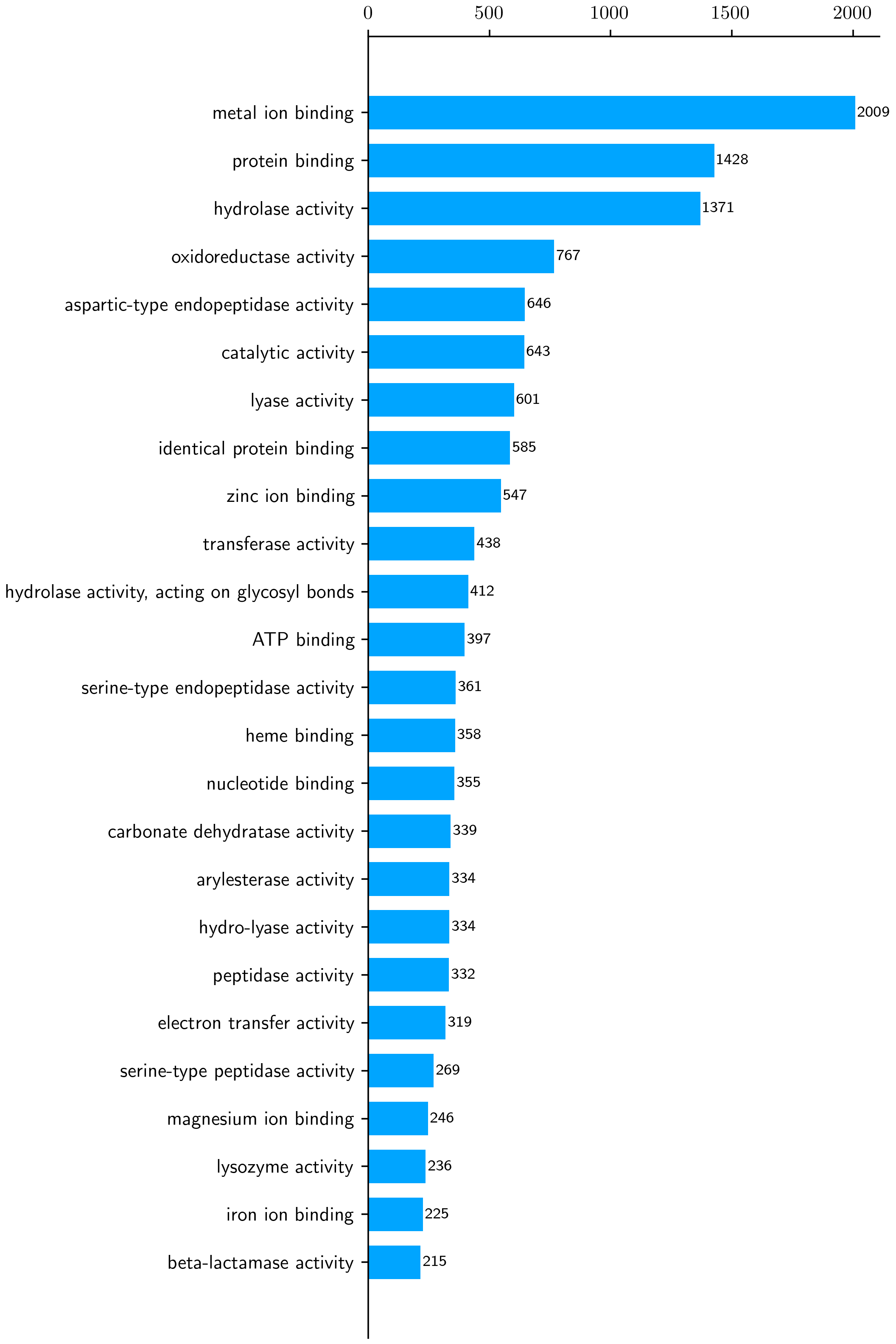}
    \caption{Summary of 25 largest MF-GO classes in our dataset.}
    \label{fig:topgoterms}
\end{figure*}

\vfill

\begin{figure*}[htb!]
    \centering
	\includegraphics[width=0.7\textwidth]{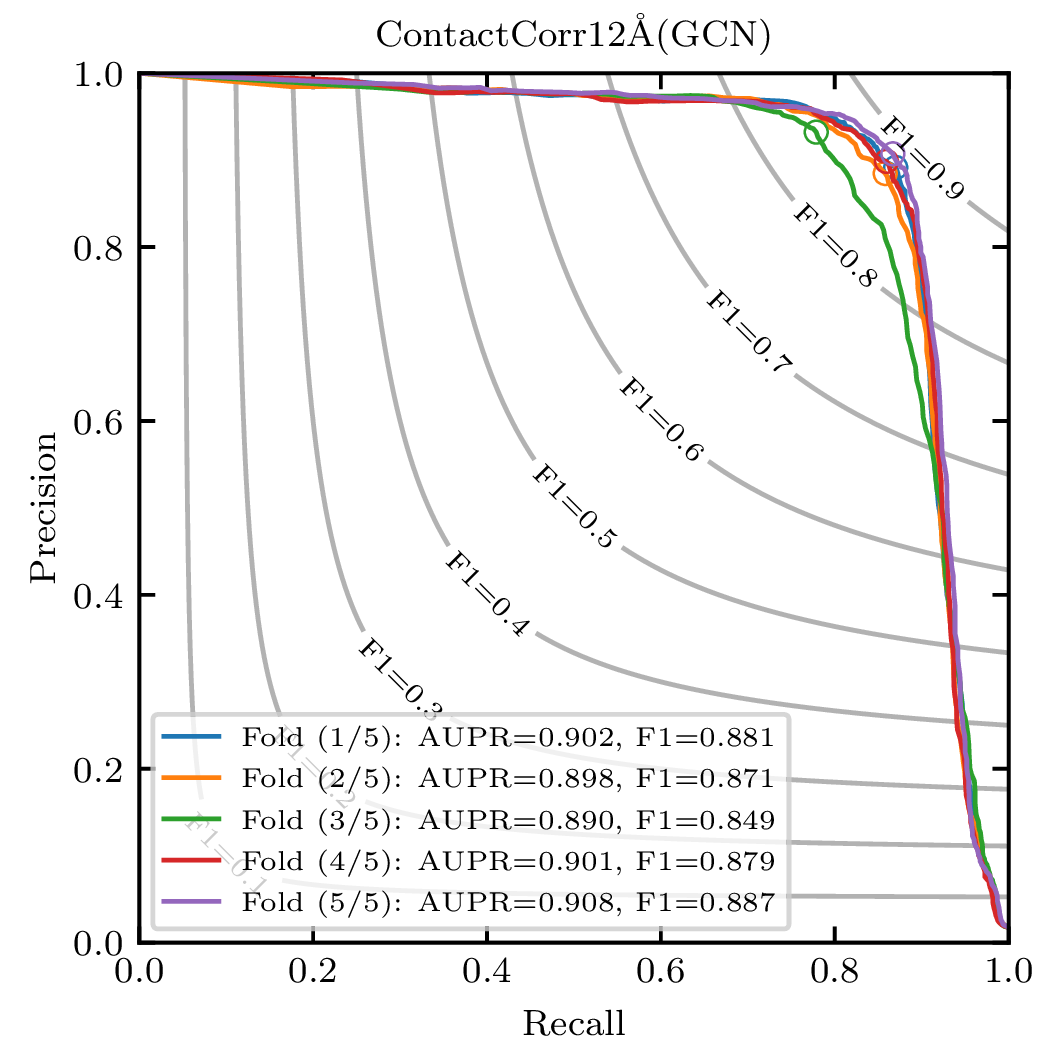}
    \caption{Five-fold cross validation of ContactCorr network with $r_c = 12$ \AA.}
    \label{fig:plot-kfold}
\end{figure*}
\clearpage

\end{document}